%% file: txhi.tex
\documentclass[12pt,a4]{article}

\usepackage{cite}
\usepackage{epsfig}
\usepackage{amsmath}
\usepackage{verbatim}
\usepackage{psfrag}
\usepackage{bm}
\usepackage{bbm}

\numberwithin{equation}{section}

\begin{document}

\begin{flushright}
{\tt MAN/HEP/2008/11}\\
July 2008
\end{flushright}

\begin{center}

{\LARGE{\bf Textures and Semi-Local Strings\\[3mm]
in SUSY Hybrid Inflation} }

\vspace{2cm}

{\large Richard A. Battye$^{\,a}$, Bj\"orn~Garbrecht$^{\,b}$ and 
Apostolos Pilaftsis$^{\,c}$}\\[0.6cm]
{\it
${}^a$Jodrell Bank Centre for Astrophysics,
School of Physics \& Astronomy,\\
University of~Manchester,
Oxford~Road,
Manchester M13~9PL, United Kingdom
}\\[3mm]
{\it ${}^b$University of Wisconsin-Madison, Department of Physics,\\
1150 University Avenue, Madison, WI 53706, USA}\\[3mm]
{\it ${}^c$School of Physics \& Astronomy,
University of~Manchester,\\
Oxford~Road,
Manchester M13~9PL, United Kingdom}

\end{center}

\vskip 1.5cm

\centerline{\bf ABSTRACT}
\medskip
\noindent
Global  topological  defects  may  account  for the  large  cold  spot
observed in the Cosmic Microwave Background.  We explore possibilities
of  constructing models of  supersymmetric $F$-term  hybrid inflation,
where the  waterfall fields  are globally ${\rm  SU}(2)$-symmetric. In
contrast  to the  case  where  ${\rm SU}(2)$  is  gauged, there  arise
Goldstone  bosons and  additional  moduli, which  are  lifted only  by
masses of  soft-supersymmetry breaking scale.  The  model predicts the
existence of  global textures, which can become  semi-local strings if
the  waterfall  fields  are  gauged under  ${\rm  U}(1)_X$.  Gravitino
overproduction  can be  avoided if  reheating proceeds  via  the light
${\rm  SU}(2)$-modes or  right-handed  sneutrinos. For  values of  the
inflaton-waterfall    coupling   $\stackrel{>}{{}_\sim}10^{-4}$,   the
symmetry breaking scale imposed by normalisation of the power spectrum
generated from  inflation coincides with the energy  scale required to
explain  the most  prominent  of the  cold  spots. In  this case,  the
spectrum of density fluctuations is close to scale-invariant which can
be reconciled with measurements of the power spectrum by the inclusion
of the sub-dominant component due to the topological defects.

\newpage

\section{Introduction}

Recent times  have seen cosmology enter a  precision era. Measurements
of the cosmic microwave background (CMB) have lead to estimates of the
standard  6 cosmological  parameters  (see, for  example,  the 5  year
results  from WMAP~\cite{Dunkley:2008ie}  and references  therein). Of
particular interest  here are the estimates of  the amplitude, $A_{\rm
s}$,  and   spectral  index,  $n_{\rm  s}$,  of   the  scalar  density
fluctuations     which     are      measured     to     be     $n_{\rm
s}=0.963^{+0.014}_{-0.015}$    and    $A_{\rm   s}=(2.41\pm0.01)\times
10^{-9}$.  It is presumed  that the  density fluctuations  are created
during an  inflationary era through quantum  effects. Constraining the
nature of this epoch on the basis of the observations is now the focus
of  much  effort~\cite{Peiris:2006sj,Kinney:2006qm,Martin:2006rs} with
the ultimate aim of making contact with particle physics models around
the Grand Unified Theory (GUT) scale~\cite{Lyth:1998xn}.

In  addition to  these fluctuations  of quantum  origin, there  may be
others  created   by  topological   defects  formed  naturally   as  a
consequence   of   a  phase   transition   at   the   end  of   hybrid
inflation~\cite{Linde:1993cn}. Such fluctuations cannot be the primary
source  of CMB  anisotropies, but  they could  provide  a sub-dominant
component~\cite{Wyman:2005tu},   which  can   have   some  interesting
effects. In particular,  it has been shown that  if cosmic strings are
formed,  then   $n_{\rm  s}=1$  can   be  made  compatible   with  the
data~\cite{Battye:2006pk,Bevis:2007gh,Battye:2007si}.    This   is  of
interest  since   many  of  the  models  of   inflation  motivated  by
fundamental theories predict larger values for the spectral index than
the WMAP best  fit value.  For example, SUSY  $F$- and $D$-term hybrid
inflationary        models~\cite{Copeland:1994vg,Dvali:1994ms,DTerminf}
predict $n_{\rm  s}>0.98$~\cite{Panagiotakopoulos:1997qd}, while even
larger values  may be required if one  considers gravitino constraints
on the reheat  temperature or the bounds on  the cosmic string tension
in  $D$-term  models.    Brane  inflation~\cite{Dvali:1998pa}  in  its
simplest form predicts $n_{\rm s}>0.97$, but again, more sophisticated
implementations    may   lead    to    a   scale-invariant    spectral
index~\cite{Haack:2008yb}.

Most work on inflation plus defect scenarios has focused on
strings. However, the recent suggestion~\cite{Cruz:2007pe,Cruz:2008sb}
that the previously identified cold spot in the CMB~\cite{Cruz:2006fy}
could be due to cosmic textures~\cite{Turok:1989ai,Turok:1990gw}
motivates our consideration of hybrid inflation scenarios involving
global defects. Such dynamically unstable defects produce a spectrum
of anisotropies qualitatively similar to strings~\cite{Bevis:2004wk}
and are likely to have similar properties in respect of the degeneracy
with $n_{\rm s}$. We note, in addition, the other cold spots detected
on smaller scale in the CMB which do not appear to be associated with
the Sunyaev-Zel'dovich effect~\cite{GenovaSantos:2008hf}.

Textures are most easily implemented in non-supersymmetric models,
where they can emerge from a scalar potential of the
form~\cite{VilenkinShellard}
\begin{equation}
\label{V:texture}
V(\Phi)=\frac14 \lambda (\Phi_i\Phi_i-\phi_0^2)^2\,,
\end{equation}
where $\lambda$  is a dimensionless coupling constant  and $\phi_0$ is
the symmetry  breaking scale.  The  fields $\Phi_i$ are  a fundamental
representation  of  a global  ${\rm  O}(4)$-symmetry.  This  potential
induces spontaneous  symmetry breaking ${\rm  O}(4)\to{\rm O}(3)$, for
which  the   vacuum  manifold   is~$S^3$,  where  $S^3$   denotes  the
3-dimensional  sphere.  Since  $\pi_3(S^3)=\mathbbm{Z}$, there  is the
possibility of non-trivial topological configurations called textures.
However, they  cannot be stable  by Derrick's theorem and  hence, once
formed at  the phase transition,  they unwind. This  unwinding process
takes  place  in a  self-similar  fashion  such  that the  density  of
textures scales relative to the  background. The latter has been shown
to be  a more  general feature  of global field  models with  a broken
$O(N)$ symmetry~\cite{Turok:1991qq}.  The process of unwinding creates
density               fluctuations               and               CMB
anisotropies~\cite{Turok:1990gw,Durrer:1990mk,Notzold:1990jt}. The CMB
anisotropies result in a distribution  of hot and cold spots, which by
the central  limit theorem, appear as approximately  Gaussian on small
scales.   On larger  scales, they  may  be revealed  by more  detailed
analysis.  By  taking into  account various observational  biases, the
analysis of  Cruz {\it  et al.}~\cite{Cruz:2007pe} has  suggested that
the large  cold spot identified  in the CMB  would result from  such a
model with
\begin{equation}
\label{eta:obs}
\phi_0=(8.7^{+2.1}_{-3.0})\times 10^{15}~{\rm GeV}\,.
\end{equation}

In this  paper, we study  supersymmetric (SUSY) scenarios  of $F$-term
hybrid inflation  that realise the global  symmetry ${\rm SU}(2)\times
{\rm  U}(1)_X$.   We also  consider  the  possibility  that the  ${\rm
U}(1)_X$-symmetry is  gauged, giving rise to a  semi-local model. Even
though  in the  context of  SUSY hybrid  inflation, models  with local
symmetries are more commonly discussed, we note that the present paper
is  not  the first  one  in  which the  use  of  global symmetries  is
considered.  In one of the  seminal papers on SUSY $F$-term inflation,
the use  of a spontaneously  broken global ${\rm  U}(1)_X$-symmetry is
suggested~\cite{Copeland:1994vg},  and  the  possibilities  of  larger
symmetries and textures  are mentioned. To the best  of our knowledge,
however,  neither an explicit  analysis of  the particle  spectrum has
been  performed  so far,  nor  have  the  consequences of  the  higher
symmetry been investigated in  detail within the framework of $F$-term
hybrid inflation.

An important  aspect of the  model-building is that  unless additional
${\rm  SU}(2)$-invariant  lifting   terms  are  included,  the  vacuum
manifold is  larger than ${\rm  SU}(2)\sim S^3$.  This is  because the
holomorphic nature of the superpotential is consistently maintained by
doubling  the number of  degrees of  freedom in  the symmetry-breaking
fields, such  as the  field $\Phi$ in~(\ref{V:texture}).   We describe
here in detail  how these additional degrees of  freedom in the vacuum
manifold  can be  lifted  such that  it  is indeed  reduced to  $S^3$.
Furthermore,  we include the  effect of  radiative corrections  on the
predictions  for  the  initial   power  spectrum  created  during  the
inflationary phase as first performed in Ref.~\cite{Dvali:1994ms}. The
breaking of a local instead of a global symmetry was first proposed in
this work  and following this  reference most articles  on SUSY-hybrid
inflation consider only local symmetries.  This may also be due to the
fact that quantum gravitational effects can violate global symmetries.
Since  no complete  model of  quantum gravity  exists yet,  we  do not
consider this possibility  here and take ${\rm SU}(2)$  to be an exact
symmetry. Such an assumption  is not entirely unrealistic, since ${\rm
SU}(2)$ is the only group that is self-non-anomalous, independently of
the representation of the chiral fermions in the theory.

The organisation of the paper  is as follows: in Section~2 we describe
in detail the  SUSY hybrid texture model based  on the global symmetry
${\rm SU(2)}\times  {\rm U(1)}_X$.  We also calculate  its scalar mass
spectrum  and discuss  mechanisms for  giving masses  to  the massless
moduli  fields   present  in  the   model.   Formal  aspects   of  the
model-building have been relegated to the appendices.  In Section~3 we
analyse  possible scenarios for  reheating after  inflation, including
constraints from  possible overproduction  of gravitinos and  Big Bang
Nucleosynthesis (BBN).  In addition, we  estimate the effect  that the
texture  Goldstone  bosons may  have  on  stellar cooling.   Section~4
discusses a semi-local realisation of the SUSY hybrid model, where the
aforementioned global  ${\rm U(1)}_X$ symmetry is promoted  to a local
one.   In Section~5  we study  the inflationary  dynamics of  the SUSY
hybrid  model and analyse  the possibility  whether the  texture scale
$\phi_0$ given in~(\ref{eta:obs})  can naturally be implemented within
this  model, without  being in  conflict with  other  cosmological and
astrophysical constraints discussed  in Section~3.  Finally, Section~6
summarises the main conclusions of our study.

\section{SUSY Hybrid Texture Model}

\subsection{Superpotential and scalar potential}

\label{section:SUSYHybrid}

The model of SUSY hybrid  inflation with global defects is implemented
by the superpotential
\begin{equation}
\label{superpotential:SU2}
W = \kappa\, \widehat{S}\, \Big( \widehat{X}_1
\widehat{X}_2\:  -\: {M}^2\Big)
\,,
\end{equation}
with $\widehat{X}_1=({\bf 2})$ being a fundamental,
$\widehat{X}_2=({\bf \overline 2})$ an identical anti-fundamental
representation of the global ${\rm SU}(2)$ and $\widehat S$ being the
singlet inflaton. We use the freedom of field redefinitions to choose
$\kappa$ and $M$ real.  In addition to the global ${\rm SU}(2)$
symmetry, the superpotential is invariant under an accidental symmetry
which we denote as ${\rm SU}(2)_X$.  In particular, ${\rm SU}(2)$
induces the transformation
\begin{equation}
\label{SU2}
\left(
\begin{array}{c}
\widehat X_1^+\\
\widehat X_1^-
\end{array}
\right)\
\xrightarrow{{\rm SU}(2)}\
{\rm e}^{{\rm i}\alpha_i\tau^i}
\left(
\begin{array}{c}
\widehat X_1^+\\
\widehat X_1^-
\end{array}
\right)
\,,\quad
\left(
\begin{array}{c}
\widehat X_2^+\\
\widehat X_2^-
\end{array}
\right)\
\xrightarrow{{\rm SU}(2)}\
{\rm e}^{-{\rm i}\alpha_i{\tau^i}^*}
\left(
\begin{array}{c}
\widehat X_2^+\\
\widehat X_2^-
\end{array}
\right)\,,
\end{equation}
while ${\rm SU}(2)_X$ has the following effect on the ${\rm
SU}(2)$-blocks:
\begin{equation}
\label{SU2X}
{\bf X}\ =\ \left(
\begin{array}{c}
\widehat X_1\\
{\rm i \tau_2}\widehat X_2
\end{array}
\right)\
\xrightarrow{{\rm SU}(2)_{\rm X}}\
{\rm e}^{{\rm i}\theta_i \sigma^i} \otimes \mathbbm{1}_2
\left(
\begin{array}{c}
\widehat X_1\\
{\rm i \tau_2}\widehat X_2
\end{array}
\right)\,.
\end{equation}
Here, $\alpha_i$ and $\theta_i$ $(i=1,2,3)$ are real numbers,
the $\tau^i$ and $\sigma^i$ denote the Pauli matrices and
$\mathbbm{1}_2$ is the two-dimensional identity matrix.
We employ the notation $\tau^i$ for
${\rm SU}(2)$-transformations and $\sigma^i$ for ${\rm SU}(2)_X$.
In the following, we assume that ${\rm SU}(2)$ is an exact symmetry,
while ${\rm SU}(2)_X$ may be explicitly broken through the introduction
of new operators in addition to those arising from the superpotential.

We  note that  the action  of a  $\theta_3$ rotation  of ${\bf  X}$ is
equivalent to an U(1) transformation given by
\begin{equation}
\widehat X_1\ \rightarrow\ {\rm e}^{{\rm i}\theta_3}\widehat X_1\,,
\quad
\widehat X_2\ \rightarrow\ {\rm e}^{-{\rm i}\theta_3}\widehat X_2\, .
\end{equation}
For  definiteness, we  identify  this particular  symmetry with  ${\rm
U}(1)_X\subset  {\rm   SU}(2)_X$,  under  which   $\widehat{X}_1$  and
$\widehat{X}_2$ transform with  opposite charges.  More formal aspects
of the  ${\rm SU}(2) \times  {\rm SU}(2)_X$ symmetry are  discussed in
Appendix~\ref{Appendix:ExtraSymmetry}.  In the main part of the paper,
we only consider the ${\rm SU}(2) \times {\rm U}(1)_X$ symmetry, since
 the  construction of  a realistic model leads to the explicit breaking
${\rm        SU}(2)_X\to{\rm       U}(1)_X$ as discussed   in
Section~\ref{sec:lifting}.

To this end,
we assume that this ${\rm SU}(2)\times{\rm U}(1)_X$
symmetry is global.
The superpotential
immediately leads to the tree-level scalar potential
\begin{eqnarray}
\label{V0}
V_0\!\!\!&=&\!\!\!\kappa^2 \left|X_1 X_2 -M^2 \right|^2\,,
\end{eqnarray}
which  is  minimised  for  $\langle  S  \rangle=0$  and  $\langle  X_1
X_2\rangle  = M^2$,  such  that  $V_0=0$. Here  and  in what  follows,
$X_1X_2$ is understood as  $X_1^{T}X_2$. The scalar fields assume this
ground-state configuration  at the  end of inflation,  while inflation
takes place for $\langle X_1 \rangle= \langle X_2\rangle=0$ and $S>M$.

Since    the    vacuum     expectation    values    (VEVs)    $\langle
X_{1,2}\rangle\not=0$ completely break the ${\rm SU}(2)$-symmetry, the
ground state is degenerate and  there will be at least 3 Goldstone
modes.\footnote{The Goldstone  modes associated with  the broken ${\rm
SU}(2)_X$ may be related to those of the ${\rm SU}(2)$-modes. For more
details,  see  Appendix~A.}  In  addition,  there  are extra  massless
modes, which we call them  moduli. These acquire masses, only when one
adds additional  ${\rm SU}(2)$-invariant  terms to the  potential that
break SUSY softly. Finally, there  are heavy modes modes of mass $\sim
\kappa {M}$.

\subsection{Identification of the scalar particle spectrum}

In order to identify these modes and determine the vacuum manifold, we
first make use  of the symmetries ${\rm SU}(2)$  and ${\rm U}(1)_X$ to
write the fields as
\begin{equation}
\label{parametrization:general}
X_1=
{\rm e}^{{\rm i}\left(\alpha_i \tau^i +\beta \mathbbm{1}_2\right)}
\left(\begin{array}{c}
{w}_1^+\\
0
\end{array}
\right)
\;
\textnormal{and}\;\;
X_2=
{\rm e}^{-{\rm i}\left(\alpha_i {\tau^i}^* +\beta \mathbbm{1}_2\right)}
\left(\begin{array}{c}
{w}_2^+ {\rm e}^{{\rm i}\varphi_2^+}\\
{w}_2^{-} 
\end{array}
\right)
\,,
\end{equation}
where $\mathbbm{1}$  is the 2-dimensional identity  matrix.  Note that
this corresponds to a completely general parameterisation of the 8 real
degrees of freedom  of $X_1$ and $X_2$ in terms  of 8 real parameters,
$w_1^+$,  $w_2^+$, $w_2^-$,  $\varphi^+_2$,  $\beta$ and  $\alpha_{i}$
$(i=1,2,3)$.

The $F$-term minimisation condition,
\begin{equation}
  \label{FScond}
F_S=\kappa\left(X_1 X_2 - M^2\right)=0\; ,
\end{equation}
then reads
\begin{equation}
{w}_1^+ {w}_2^+ {\rm e}^{{\rm i}\varphi_2^+}=M^2\; .
\end{equation}
This  last  equality  immediately  implies that  $\varphi_2^+=0$.  The
minimisation  condition~(\ref{FScond})   will  persist  in   the  more
complicated  situations considered in  our subsequent  discussion.  We
therefore define
\begin{equation}
{w}=\sqrt{{{w}_1^+}^2+{{w}_2^+}^2+{{w}_2^-}^2}\; ,
\end{equation}
and note the relation
\begin{equation}
\label{rho>sqrt2*M}
{w}\geq \sqrt{{{w}_1^+}^2+{{w}_2^+}^2}\geq \sqrt 2 M\,,
\end{equation}
which    has   to    hold    true   in    order    to   satisfy    the
condition~(\ref{FScond}) for the $F_S$ term.

In   order   to   determine   the   mass  eigenmodes,   we   use   the
parameterisation~(\ref{parametrization:general}) and,  without loss of
generality  due to  the global  symmetries, pick  a  convenient point,
e.g.~$\alpha_i=\beta=0$,  to  expand  about   it.  In  this  way,  the
waterfall doublets $X_{1,2}$ may be expressed as
\begin{equation}
 \label{unitarygauge}
X_{1}=
\left(
\begin{array}{c}
{w}_1^+ +\frac{1}{\sqrt2}(X_{1R}^+ +{\rm i}X_{1I}^+)\\
\frac{1}{\sqrt2}(X_{1R}^- +{\rm i}X_{1I}^-)
\end{array}
\right)\;
\textnormal{and}\;\;
X_{2}=
\left(
\begin{array}{c}
{w}_2^+
+\frac{1}{\sqrt2}(X_{2R}^+ +{\rm i}X_{2I}^+)\\
{w}_2^- + \frac{1}{\sqrt2}(X_{2R}^- +{\rm i}X_{2I}^-)
\end{array}
\right)\,,
\end{equation}
where ${w}^\pm_{1,2}$ denote the classical VEVs and $X^\pm_{1,2\,R,I}$
are the respective quantum excitations.

The massless ${\rm SU}(2)$ Goldstone modes are
\begin{subequations}
\label{Goldstones}
\begin{eqnarray}
{\rm SU}(2)&:&\;G_{\tau^1}=
\frac1{w}\left(
{w}_1^+ X_{1I}^- - {w}_2^+ X_{2I}^- -{w}_2^- X_{2I}^+
\right)\,,\\
&&\;G_{\tau^2}=
\frac1{w}\left(
-{w}_1^+ X_{1R}^- - {w}_2^+ X_{2R}^- + {w}_2^- X_{2R}^+
\right)\,,\\
&&\;G_{\tau^3}=
\frac1{w}\left(
{w}_1^+ X_{1I}^+ - {w}_2^+ X_{2I}^+ + {w}_2^- X_{2I}^-
\right)\,.
\end{eqnarray}
\end{subequations}
The  indices attached  to $G$  indicate the  particular  generators of
${\rm  SU}(2)$ that  these modes  correspond to.   For  the subsequent
discussion, we also note the massless ${\rm U}(1)_X$-direction
\begin{equation}
\label{U1Goldstone}
{\rm U}(1)_X :\;G_{\mathbbm{1}}=
\frac1{w}\left(
{w}_1^+ X_{1I}^+ - {w}_2^+ X_{2I}^+ -{w}_2^- X_{2I}^-
\right)\,.
\end{equation}
However,  it  is  not  orthogonal  to  the  ${\rm  SU}(2)$-modes  and,
therefore,  we do not  use it  here as  a local  basis vector  for the
scalar    field    space.    Moreover,    for   $w_2^-=0$,    it    is
$G_{\mathbbm{1}}=G_{\tau^3}$.

In  addition to the  Goldstone bosons,  there are  massless eigenmodes
which  are not  protected  by  the ${\rm  SU}(2)$  and ${\rm  U}(1)_X$
symmetries.  We  refer to them  as moduli, and  note that they  may be
lifted    by   adding    additional   ${\rm    SU}(2)$-    and   ${\rm
U}(1)_X$-invariant  terms to the  scalar potential  or by  gauging the
${\rm U}(1)_X$.  The modes of this type are
\begin{subequations}
\begin{eqnarray}
H_0^R\!\!\!&=&\!\!\!\frac 1{w} \left({w}_2^+ X_{1R}^- -
{w}_1^+ X_{2R}^- - {w}_2^- X_{1R}^+
\right)\,,\\
H_0^I\!\!\!&=&\!\!\!\frac1{{w}}\left({w}_2^+ X_{1I}^- + {w}_1^+ X_{2I}^-
-{w}_2^- X_{1I}^+ \right)\,,
\end{eqnarray}
\end{subequations}
and
\begin{equation}
\label{Hg}
H_g=
\frac1{w}\left(
-{w}_1^+ X_{1R}^+ + {w}_2^+ X_{2R}^+ -{w}_2^- X_{2R}^-
\right)\,.
\end{equation}
As well as the massless  fields, the scalar particle spectrum contains
massive eigenmodes. These are
\begin{eqnarray}
  \label{HRkappa}
H^R_{\kappa} \!\!\!&=&\!\!\!
\frac1{w}\left(
{w}_1^+ X_{2R}^+ + {w}_2^+ X_{1R}^+ +{w}_2^- X_{1R}^-
\right)\,,\quad
H^I_{\kappa} = \frac1{w}\left(
{w}_1^+ X_{2I}^+ + {w}_2^+ X_{1I}^+ +{w}_2^- X_{1I}^-
\right)\,,\qquad\\
S^R \!\!\!&=&\!\!\! {\rm Re}\,S\, ,\quad S^I\ =\ {\rm Im}\,S\; . 
\end{eqnarray}
All  these  states are  degenerate  with  mass  equal to  $\kappa\,w$.
Because of  this degeneracy, the mass eigenstates  may not necessarily
be orthogonal to each other. However, our field basis,
\begin{equation}
\label{orthoscalar}
\left\{G_{\tau^1},\, G_{\tau^2},\, G_{\tau^3},\, H_0^R,\, H_0^I,\, H_g,\,
H_\kappa^R,\, H_\kappa^I,\, S^R,\, S^I \right\}\,,
\end{equation}
has been so chosen to be orthonormal.

\subsection{Lifting of the moduli}
\label{sec:lifting}

Unlike  the Goldstone  bosons~(\ref{Goldstones}),  the moduli  fields,
$H_0^R$, $H_0^I$ and  $H_g$, are not protected by  the global symmetry
${\rm SU}(2)\times{\rm U}(1)_X$ and usually acquire non-zero masses if
soft SUSY-breaking terms are added  to the potential which respect the
global symmetry.  We refer to  this mechanism as lifting.  Some of the
lifted  modes turn  out to  correspond to  broken generators  of ${\rm
SU}(2)_X$, which  is explicitly  broken by the  lifting mass  terms. A
categorisation     of     these      modes     is     presented     in
Appendix~\ref{Appendix:ExtraSymmetry}.

The possibility of adding terms to the superpotential of the type
\begin{equation}
\label{lifting:possibly}
\widehat X_{1,2} \widehat T \widehat X_{1,2}\,,\quad
\widehat X_{1,2} \widehat T \widehat X_{2,1}\,,\quad
\mu_X{\widehat X_1}{\widehat X_2}\,,
\end{equation}
where $\widehat T$  denotes an ${\rm SU}(2)$-triplet and  $\mu_X$ is a
mass of  order $M$  has to be  discarded. The $\mu_X$-term  breaks the
$R$-symmetry    of    the    superpotential~(\ref{superpotential:SU2})
explicitly, such that there is no SUSY-preserving minimum. The triplet
variant  leads  to spontaneous  $R$-symmetry  breaking  and again,  no
supersymmetric minimum occurs~\cite{Nelson:1993nf}.

The generic form for a potential representing soft SUSY breaking in
the inflaton-waterfall sector is
\begin{equation}
\label{Vsoft}
V_{\rm soft}=
m_1^2 |X_1|^2 + m_2^2 |X_2|^2 + b X_1^T X_2 + b^* X_1^\dagger X_2^*
+d X_1^\dagger {\rm i} \tau^2 X_2 - d^* X_2^\dagger {\rm i}\tau^2 X_1\,.
\end{equation}
We now briefly discuss the possible  origin of the 3 types of mass
terms in~(\ref{Vsoft}) and their physical significance:
\begin{itemize}

\item[(i)]  The terms  $m_{1,2}$ may  directly result  from  soft SUSY
breaking. Alternatively, they could arise by giving a TeV-scale VEV to
$S$,  as discussed  in Refs.~\cite{Dvali:1997uq,King:1997ia,Garbrecht:2006az},
which would also address the $\mu$-problem of the
Minimal Supersymmetric Standard Model (MSSM).

\item[(ii)] The  $b$ term may  be a soft SUSY-breaking  parameter that
could    originate     from    a    superpotential     term,    $\mu_X
\widehat{X}_1\widehat{X}_2$,  as  given  in  (\ref{lifting:possibly}),
where $\mu_X$  is of the soft SUSY-breaking  scale~$M_{\rm SUSY}$.  If
$b  \sim  M^2_{\rm  SUSY}$,  the  $b$-terms  lead  to  a  sub-dominant
correction   to    the   $F_S=0$   condition    for   minimising   the
potential. Hence, they are not  suitable for lifting the moduli and we
do not discuss them here any further.

\item[(iii)] The  $d$-terms may occur  as a result of  a non-canonical
K\"ahler potential.   Although they break the  accidental global ${\rm
U}(1)_X$-symmetry, we  show in Appendix~\ref{Appendix:FieldRedef}, how
these $d$-terms can  be absorbed in $m_1^2$ and  $m_2^2$ after a field
redefinition of $X_1$ and $X_2$.
 
\end{itemize}

Consequently,  we will  set both  $b=0$  and $d=0$  in the  subsequent
discussion and just allow $m_{1,2}^2$ to vary, that is
\begin{equation}
\label{Vlift}
V_{\rm lift} = m_1^2 |X_1|^2 + m_2^2 |X_2|^2\; .
\end{equation}
Finally, to  keep things  at a  more general level,  we prefer  not to
specify  the soft SUSY-breaking  masses for  the superpartners  of the
Goldstone bosons and the light  moduli.  For instance, we could assume
that these particles are very light and decouple above the electroweak
symmetry  breaking   scale,  such  that   they  constitute  additional
relativistic species today (see also our discussion in Section~3.4).

\subsection{Vacuum manifold and texture defects}

We now determine the vacuum manifold of the scalar potential,
which arises as a combination of~(\ref{V0}) and~(\ref{Vlift}),
\begin{equation}
V_{\rm global}=V_0+V_{\rm lift}\,.
\end{equation}
With the parameterisation~(\ref{unitarygauge}),
we get the potential
\begin{equation}
V_{\rm global}({w}_1^+,{w}_2^+,{w}_2^-,\varphi_2^+)=
\kappa^2\left({w}_1^+{w}_2^+e^{i\varphi_2^+} - {M}^2 \right)^2
+m_1^2{{w}_1^+}^2+m_2^2\left({{w}_2^+}^2+{{w}_2^-}^2\right)\,.
\end{equation}
We find its minimum at
\begin{eqnarray}
\label{dda}
w_1^+=\sqrt{{m_2\over m_1}M^2-{m_2^2\over\kappa}}\,,\quad 
w_2^+=\sqrt{{m_1\over m_2}M^2-{m_1^2\over\kappa}}\,,\quad
w_2^{-}=0\,,\quad\varphi_2^+=0\,.
\end{eqnarray}
At this point, the moduli are lifted to acquire the masses
\begin{eqnarray}
\label{moduli:masses}
m^2(H_0^R)=m^2(H_0^I)=m_1^2+m_2^2\,,\qquad  m^2(H_g)\!\!\!&=&\!\!\! 4
\frac{m_1^2 m_2^2}{m_1^2 + m_2^2}\,. 
\end{eqnarray}

Recalling that  the condition  $w_2^-=0$ implies a  degeneracy between
the     parameters      $\alpha_3$     and     $\beta$      in     the
parameterisation~(\ref{parametrization:general}),                    the
minimum~(\ref{dda}) fixes 5 out of 8 degrees of freedom.  Hence, after
the  fixing of the  moduli, the  vacuum manifold  is $S^3$,  where the
sphere   is    generated   by   the   3    Goldstone   modes   defined
in~(\ref{Goldstones}).  In this model, there are therefore textures of
a     symmetry     breaking      scale     $\sqrt     2     M$.     In
Appendix~\ref{Appendix:ExtraSymmetry}, we present  a discussion of the
vacuum manifold in the case where $V_{\rm lift}=0$.

\section{Cosmological and Astrophysical Implications}
\label{section:X:MSSM}

In  this section, we  outline possible  scenarios for  reheating after
inflation. In order  to do so, we choose to  consider couplings to the
Higgs-fields of  the MSSM and  the right-handed neutrino  sector, such
that  the  inflationary  energy   can  eventually  be  converted  into
particles  of the Standard  Model (SM).  Furthermore, we  estimate the
effect the new  light degrees of freedom in  the waterfall sector have
on  stellar  cooling  and   on  the  production  of  gravitinos  after
inflation.  We also note that  the possible existence of the new light
fields  is not  in  conflict with  bounds  on additional  relativistic
degrees of freedom that arise from BBN.

\subsection{Extended superpotential}

We extend the superpotential~(\ref{superpotential:SU2}) by adding the
terms
\begin{equation}
W_{\rm RH} = \lambda\, \widehat{S} \widehat{H}_u
\widehat{H}_d\ +\ \frac{\rho_{ij}}{2} \widehat{S}\, \widehat{N}_i
\widehat{N}_j\,, 
\end{equation}
where $\widehat H_{u,d}$ denote  the MSSM-Higgs doublets and $\widehat
N_i$ ($i=1,2,3$)  are right-handed singlet neutrinos,  which we assume
to        be        nearly        degenerate,        $\rho_{ij}\approx
\rho\,\mathbbm{1}_3$~\cite{Pilaftsis:2004xx,Pilaftsis:2005rv}.  Again,
we  use the  freedom of  field redefinitions  to choose  $\lambda$ and
$\rho$  to be  real.  The  tree-level scalar  potential~(\ref{V0}) now
generalises to
\begin{eqnarray}
\label{V0:lambda:rho}
V_0\!\!\!&=&\!\!\!\kappa^2 \left|X_1X_2 + \frac{\lambda}{\kappa} H_u H_d
+\frac{\rho_{ij}}{2 \kappa} N_i N_j - {M}^2 \right|^2
+\kappa^2 |S X_1|^2 + \kappa^2 |S X_2|^2
\\
\nonumber
&&
+ \lambda^2|S H_u|^2 + \lambda^2|S H_d|^2 +
|\rho_{ij} S N_j|^2\,.
\end{eqnarray}
In Appendix C, we present the terms that are quadratic in $X_{1,2}$,
such that possible interactions within this sector and with
the Higgs fields and right-handed neutrinos can easily be read off.

\subsection{Stellar cooling}

It is  straightforward to check  that the coupling $\lambda$  does not
induce any  tree-level mixing between the Higgs  bosons, the Goldstone
bosons  and  moduli.   Hence,   we  do  not  expect  related  collider
phenomenology.

We may now examine whether  the same coupling induces effects that are
in contradiction with astrophysical  observations, especially due to a
dangerous increase in the rate of stellar energy loss. To this end, we
first observe  that because $\widehat  X_{1,2}$ are not  charged under
hypercharge  or weak  isospin, so  there  is no  mixing between  these
fields and the  SM gauge bosons.  From ~(\ref{X1X2}),  we see that the
light  moduli  and Goldstone  fields  are  always  produced in  pairs.
Hence,  possible Bremsstrahlung  processes,  where a  fermion of  mass
$m_f$ emits  two Goldstone bosons via a  virtual Higgs-boson exchange,
involve a factor
\begin{equation}
{\kappa\lambda m_f T_{\rm S}\over 2\pi^2m_H^2}\,,
\end{equation}
when compared  to the axion-Yukawa  coupling $\sim m_f/f_a$,  which is
relevant   for   the  process   of   axion  Bremsstrahlung,   e.g.~see
Ref.~\cite{Raffelt:1990yz}.  Here,  $T_{\rm S}$ is  the temperature of
the  star,  $m_H$  the mass  of  the  SM  Higgs  boson and  $f_a$  the
axion-decay      constant.      With      the      supernova     bound
$f_a\stackrel{>}{{}_\sim}10^{10}~{\rm     GeV}$,     $T_{\rm    S}\sim
10^{-1}~{\rm GeV}$ in  the core of a supernova  and $m_H\sim 10^2~{\rm
GeV}$, it  is clear that  fast energy loss  can be avoided  if $\kappa
\lambda  \stackrel{<}{{}_\sim}10^{-4}$, which  can easily  be achieved
for      the      values       of      $\kappa$      discussed      in
Section~\ref{section:inflation}.

\subsection{Reheating and gravitinos}

We sketch  two possible ways  in which the inflationary  vacuum energy
can   be  transferred  to   the  MSSM   degrees  of   freedom  through
renormalisable   interactions.    One   may   alternatively   consider
``gravity-mediated" reheating  through non-renormalisable operators, a
possibility which we do not discuss here.

For  the first  reheat scenario,  we assume  that  $\lambda\not=0$ and
$\rho=0$.  In order to  avoid gravitino  overproduction mediated  by a
fast  decay of  the inflaton  into the  MSSM sector,  a large  VEV for
$\widetilde t_R$ or $\widetilde  t_L$ is required during inflation and
the  initial stages  of reheating.  The MSSM  degrees of  freedom only
equilibrate after relaxation of the  stop VEV.  In the second case, we
set  $\lambda=0$ and  $\rho>\kappa$, such  that the  $\widehat  N$ can
mediate between the $\widehat  X_{1,2}$ and the MSSM. Equilibration of
the   MSSM  degrees  of   freedom  is   then  delayed   through  small
neutrino-Yukawa couplings.

\subsubsection{Reheating delayed by stop VEV}
\label{stop:VEV:delay}

Considering  the  scalar  potential~(\ref{V0})  and  the  interactions
arising from~(\ref{X1X2}), it is clear that the oscillating fields $S$
and  $H_\kappa^R$  can directly  decay  into  Goldstone bosons,  light
moduli and their superpartner fermions. Since these particles are only
weakly  coupled to the  MSSM degrees  of freedom,  higher temperatures
within this sector  are in accordance with gravitino  bounds.  This is
because gravitino production from  MSSM degrees of freedom is mediated
by non-Abelian and super-gauge  interactions, which are absent for the
fields $\widehat  X_{1,2}$. We  now estimate the  gravitino abundance,
which  arises from  a thermal  bath of  the light  degrees  of freedom
within these fields.

The decay rate of the oscillating fields is
\begin{equation}
\Gamma_{\kappa}=\frac{3}{32\pi}
\kappa^3{w}\,,
\end{equation}
and the reheat temperature within the light components of $X_{1,2}$
\begin{equation}
T_{\kappa}=
\left(\frac{45}{4 \pi^3 g_*^{\rm G}}\right)^{1/4}
\sqrt{\Gamma_{\kappa} m_{\rm Pl}}
=2.1 \times 10^{16}~{\rm GeV}\, \kappa^{3/2}
\sqrt{\frac{{w}}{10^{16}~{\rm GeV}}}
\,,
\end{equation}
where the effective
number of degrees of freedom within the Goldstone bosons,
light moduli and their superpartners is given by
$g_*^{\rm G}=11.25$.
Considering the scalar interactions arising from~(\ref{X1X2})
in combination with the interaction of a gravitino and a chiral multiplet
as given for example in Ref.~\cite{Kawasaki:1994af}, we see that gravitinos
$\widetilde G$ may be produced in $2\to 3$ processes of the type
\begin{equation}
Y + Y \to \widetilde G + Y + \widetilde Y\,,
\end{equation}
where $Y$ stands for any of the Goldstone bosons or moduli.  There are
also $3 \to 2$ processes of equal importance, which are related to the
$2\to 3$ by crossing. We estimate the averaged cross section for the
$2\to 3$ as
\begin{equation}
\langle \sigma v \rangle
\simeq
\frac{6 \kappa^2}{8 m_{\rm Pl}^2 (2\pi)^2}\,.
\end{equation}
Compared   to   the   expressions   for   a  $2\to   2$   process   in
Ref.~\cite{Kawasaki:1994af}, we have inserted  a phase space factor of
$(2\pi)^{-2}$, as  well as a factor  of 6, which counts  the number of
light  chiral  multiplets   arising  from  $\widehat{X}_{1,2}$.   This
estimate should be accurate up to  a factor of order 1.  The gravitino
number can then be calculated as~\cite{Kawasaki:1994af}
\begin{equation}
Y_{\widetilde G}=\frac{g_*^{1 {\rm MeV}}}{g_*^{\rm G}}
\frac{n(T_\kappa) \langle \sigma v \rangle}{H(T_\kappa)}
=8\times 10^{-17}\left(\frac{\kappa}{10^{-3}}\right)^{7/2}
\sqrt{\frac{{w}}{10^{16}~{\rm GeV}}}\ ,
\end{equation}
where  $n(T_\kappa)=\zeta(3)T^3/\pi^2$ and  $g_*^{1  {\rm MeV}}=3.91$.
This   typically  is   below  the   upper  bounds   $Y_{\widetilde  G}
\stackrel{<}{{}_\sim}10^{-15}$ for a  gravitino of mass $m_{\widetilde
G}\simeq      360~{\rm      GeV}$      and      $Y_{\widetilde      G}
\stackrel{<}{{}_\sim}10^{-14}$ for a  gravitino of mass $m_{\widetilde
G}\simeq   600~{\rm   GeV}$,    as   imposed   by   constraints   from
BBN~\cite{Kawasaki:2004yh,Kawasaki:2004qu}.   Larger  gravitino masses
relax  this bound. Therefore,  the Goldstone  and light  moduli sector
does not lead to a harmful amount of gravitino production.

If we alternatively assume that $\lambda\not=0$ and if the $\widehat
H_{u,d}$ are massless at reheating, then the inflaton can also
directly decay into Higgs bosons and Higgsinos.  The condition
$\lambda > \kappa$ then leads to a reheat temperature of
\begin{equation}
T_{\rm R}=\left(\frac{45}{4\pi^3 g_*}\right)^{1/4}
\left(
\frac{m_{\rm Pl}}{32 \pi} 4\lambda^2 \kappa {w}
\right)^{1/2}
=7\times 10^{15}~{\rm GeV} \lambda \sqrt{\kappa}
\left(\frac{{w}}{10^{16}~{\rm GeV}}\right)^{1/2}\,,
\end{equation}
within the MSSM sector, where $g_*=240$ for the MSSM with right-handed
neutrinos. Unless  $\kappa$ and  $\lambda$ are tuned  to be  less than
about $10^{-5}$, this gives rise to unacceptably high temperatures. On
the other hand, these small  values would suppress the energy scale of
the texture  way below the value  reported in Ref.~\cite{Cruz:2007pe},
as  we  see in  Section~\ref{section:inflation}.   Decays  of $S$  and
$H^{R,I}_\kappa$  into $\widehat H_{u,d}$  can, however,  initially be
avoided if  the MSSM-Higgs  particles acquire a  large mass  through a
large VEV along  a flat direction involving the  left- or right-handed
stop, such that $\kappa{M} \ll \langle \widetilde t \rangle$.

For a large stop VEV, we  can also allow for $\lambda < \kappa$, since
the $H_{u,d}$  mass eigenvalues  remain positive during  inflation. To
see  this, let  us  denote the  stop  VEV as  $v_{\rm  stop}$ and  the
top-Yukawa     coupling    as    $h$.      In    the     weak    basis
$\left(H_u^*,H_d\right)$,  the Higgs  mass-square matrix  may  then be
written down as
\begin{equation}
\left(
\begin{array}{cc}
\lambda^2|S|^2+h^2 v_{\rm stop}^2 & -\kappa\lambda{M}^2\\
-\kappa\lambda {M}^2 & \lambda^2 |S|^2
\end{array}
\right)\,,
\end{equation}
and the condition for it to be positive definite is
\begin{equation}
  \label{Scond}
|S|^2 >
\frac{-h^2 v_{\rm stop}^2+\sqrt{4\kappa^2\lambda^2 {M}^4+h^4 v_{\rm stop}^4}}
{2\lambda^2}
\approx
\frac{\kappa^2{M}^4}{h^2 v_{\rm stop}^2}\,,
\end{equation}
where  we have assumed  $h^2 v_{\rm  stop}^2\gg4\kappa\lambda{M}^2$ to
arrive  at the  last expression  in~(\ref{Scond}).   Therefore, during
inflation,  where  $\langle|S|\rangle>M$,   the  Higgs  masses  remain
positive definite.

If the flat  direction is lifted by an operator of  TeV scale, its VEV
will relax towards zero  at temperatures of about $10^{10}~{\rm GeV}$.
Another possibility  is the fragmentation  of the flat  direction into
Q-balls~\cite{Berkooz:2005sf,Berkooz:2005rn}.  To that end, all energy
is contained within the $\widehat X_{1,2}$. For the scatterings of the
$X_{1,2}$ into $H_{u,d}$, we estimate $\langle \sigma v\rangle n\simeq
\kappa^2 \lambda^2 T_{\rm R}$.   The reheat temperature $T_{\rm R}$, at
which the MSSM degrees of freedom equilibrate, can be estimated as
\begin{equation}
\langle \sigma v \rangle n \simeq H \simeq \frac{T_{\rm R}^2}{m_{\rm
    Pl}}\, .
\end{equation}
Thus, up to factors ${\cal O}(1)$, we find that
\begin{equation}
T_{\rm R}\simeq 5 \times 10^{5}~{\rm GeV}
\left(\frac{\kappa}{10^{-3}}\right)^2
\left(\frac{\lambda}{10^{-3}}\right)^2\,.
\end{equation}
Reheat  temperatures   which  are  not  in   conflict  with  gravitino
overproduction can therefore easily be attained for values of $\kappa$
which are  suggested by the scale  of the texture, which  we derive in
Section~\ref{section:inflation}.

\subsubsection{Reheating via sneutrinos}

If  $\lambda=0$  and $\rho  \geq  \kappa$,  decays  into right  handed
neutrinos and sneutrinos also take place.  The decay rate is
\begin{equation}
\Gamma_{\kappa\rho}=\frac{3}{32\pi}
\left(\kappa^2+\rho^2\right)\kappa{w}\,,
\end{equation}
and the reheat temperature within this sector is given by
\begin{equation}
T_{\kappa\rho}=
\left(\frac{45}{4 \pi^3 g_*^{\rm G}}\right)^{1/4}
\sqrt{\Gamma_{\kappa_\rho} m_{\rm Pl}}
=1.2 \times 10^{16}~{\rm GeV} \sqrt{3\kappa(\kappa^2+\rho^2)}
\sqrt{\frac{{w}}{10^{16}~{\rm GeV}}}
\,,
\end{equation}
where the effective number of  degrees of freedom within the Goldstone
bosons, light  moduli, right handed neutrinos  and their superpartners
is given by  $g_*^{\rm G}=22.5$. By the same  argument as given above,
there is no gravitino overproduction within this sector.

Potentially  dangerous   interactions  with  gravitinos   only  become
important  when  MSSM  degrees  of  freedom,  which  have  strong  and
electroweak interactions,  are produced  in sizeable amounts.   Let us
therefore  assume  that the  largest  neutrino  Yukawa  coupling in  a
diagonal basis is given by $h$ and add the superpotential term
\begin{equation}
h \widehat L \widehat H_u N\,.
\end{equation}
Then,  according  to  the  standard  seesaw  mechanism,  the  heaviest
neutrino mass-eigenstate is of mass $m_\nu\simeq h^2 v^2 / m_N$, where
$v=\langle H_u^0\rangle$ and $m_N$  the mass of the heavy right-handed
neutrinos $N_i$, which we assume to  be of TeV scale. In this picture,
all the  neutrino-Yukawa couplings $h$ are  ${\cal O}(10^{-6})$, which
may be regarded as somewhat unnatural.

The scattering rate of the right-handed neutrinos with left-handed
neutrinos and Higgs particles or respective superpartners is given
by the averaged cross section $\langle \sigma v \rangle n$.
The reheat temperature $T_{\rm R}$, at which the MSSM degrees of freedom
equilibrate, can be estimated as
\begin{equation}
\langle \sigma v \rangle n\simeq h^2 T_{\rm R} \simeq H \simeq
\frac{T_{\rm R}^2}{m_{\rm Pl}}\,, 
\end{equation}
up to  factors ${\cal O}(1)$.  For an order-of-magnitude  estimate, we
find
\begin{equation}
T_{\rm R}\simeq 10^5~{\rm GeV} \left(\frac{m_N}{1 {\rm TeV}}\right)
\left(\frac{m_\nu}{50 {\rm meV}}\right) 
\left(\frac{200~{\rm GeV}}{v}\right)^2\,.
\end{equation}
Reheat temperatures of  the MSSM which are low  enough not to conflict
with  the gravitino  bound  but  high enough  to  allow for  low-scale
resonant    leptogenesis~\cite{Pilaftsis:1997jf,Pilaftsis:2003gt}   or
electroweak  baryogenesis~\cite{EWBAU}  can  be  achieved  within  the
suggested model.

\subsection{Big bang nucleosynthesis and the new light fields}

The number and the temperature of light relativistic species is
constrained by considerations from BBN, or even
more strongly by a combined analysis of CMB data and observations
of abundances of light nuclei, see for example
Ref.~\cite{Barger:2003zg}. This is because the results of BBN
are sensitive to the effective number of degrees of
freedom. We denote by $N_\nu$ the number of neutrinos, which freeze
out before electron-positron annihilation but after the annihilation
of any other species, where $N_\nu=3$ in the MSSM and its extension
which we discuss in this paper.  We consider the case when reheating
takes place according to the stop-VEV scenario, as outlined in
Section~\ref{stop:VEV:delay}; the neutrino-reheating scenario will be
very similar.

The  Goldstone bosons  $G_{\tau^i}$ will  freeze out,  as soon  as the
temperature of the Universe drops below the mass of the lightest Higgs
boson  or  Higgsino.   If  we  assume that  all  other  supersymmetric
particles and the top-quark are  heavier than the lightest Higgs boson
or  Higgsino, we  can  estimate  the effective  number  of degrees  of
freedom  at this point  to be  $g_*^{\rm freeze-out}\simeq  100$.  For
definiteness,  besides the  Goldstone bosons,  we assume  that  also 6
Weyl-fermions  composed  from  $\widetilde X_{1,2}$  are  relativistic
during BBN. The  effective number of degrees of  freedom today is then
given by
\begin{equation}
g_*=2+\frac 78 \times 2\times N_\nu
\left(
\frac{4}{11}
\right)^{4/3}
+\left[
\left(3+6\times \frac 78\right)
\left(
\frac{2}{g_*^{\rm freeze-out}}
\right)^{4/3}
\right]\,.
\end{equation}
Numerically, the last  term is $0.04$, while the sum  of the first two
terms is 3.4. Since the  present bounds on $N_\nu$ are $1.7$--$3.2$ at
$3\sigma$ confidence level~\cite{Barger:2003zg}, the contribution from
the new  light particles of our  model to $g_*$ does  not endanger the
successful predictions of BBN.

\section{Semi-Local Model}

It is possible  to promote the global symmetry to  a semi-local one by
assigning opposite  ${\rm U}(1)_X$ gauge  charges to $X_1$  and $X_2$.
In this case, the scalar-gauge Lagrangian is
\begin{equation}
\label{Lagrangian:U1}
{\cal L}=-\frac 14 F_{\mu\nu}F^{\mu\nu}
-|D_\mu X_1|^2-|D_\mu X_2|^2-V_0- V_{\rm lift} -\frac 12 D^2\,.
\end{equation}
For the kinetic terms, we use the standard abbreviations
\begin{equation}
D_\mu X_{1,2}=\partial_\mu X_{1,2} \pm{\rm i} \frac g2 A_\mu X_{1,2}
\end{equation}
and
\begin{equation}
F_{\mu\nu}=\partial_\mu A_\nu - \partial_\nu A_\mu\,.
\end{equation}
The $D$-term is given by
\begin{equation}
\label{DTerm}
D=\frac g2 (X_1^*X_1 - X_2^* X_2 -m_{\rm FI}^2)\, .
\end{equation}
Evidently, the  global ${\rm SU}(2)$ model  is the $g=0$  limit of the
semi-local one.  Note  that if we do not gauge  the ${\rm U}(1)_X$, it
still       is       a        global       symmetry       of       the
superpotential~(\ref{superpotential:SU2}).   We  also   allow   for  a
Fayet-Iliopoulos mass term $m_{\rm FI}$, which we assume to be $m_{\rm
FI}\ll{M}$.

Here,  we only  discuss  the  most important  changes  that the  ${\rm
U}(1)_X$-gauge interaction  incurs when compared to  the purely global
case.   Choosing again  the  parameterisation~(\ref{unitarygauge}) and
using $V_{\rm lift}$, we minimise
\begin{eqnarray}
V_{{w}}({w}_1^+,{w}_2^+,{w}_2^-,\varphi_2^+)&=&
\kappa^2\left({w}_1^+{w}_2^+e^{i\varphi_2^+} - {M}^2 \right)^2
+m_1^2{{w}_1^+}^2+m_2^2{{w}_2^+}^2+m_2^2{{w}_2^-}^2\cr
&+&\frac{g^2}{8}\left({{w}_1^+}^2-{{w}_2^+}^2-{{w}_2^-}^2-m_{\rm
  FI}^2\right)^2\,, 
\end{eqnarray}
and find, assuming $m_{1,2}\ll{M}$,
\begin{subequations}
\label{VEV:semilocal}
\begin{eqnarray}
{w}_1^+\!\!\!&=&\!\!\!
{M}-\frac{2\kappa^2(m_1^2-m_2^2-\frac{g^2}{2}m_{\rm
    FI}^2)+g^2(m_1^2+m_2^2)}{4 g^2 \kappa^2 {M}} 
+O\left(\frac{m_{1,2}^4}{{M}^3}\,,\frac{m_{\rm FI}^4}{{M}^3}\right)\,,
\\
{w}_2^+\!\!\!&=&\!\!\!
{M}-\frac{2\kappa^2(m_2^2-m_1^2+\frac{g^2}{2}m_{\rm
    FI}^2)+g^2(m_1^2+m_2^2)}{4 g^2 \kappa^2{M}} 
+O\left(\frac{m_{1,2}^4}{{M}^3}\,,\frac{m_{\rm FI}^4}{{M}^3}\right)\,,
\\
\label{w2:semilocal:0}
{w}_2^-\!\!\!&=&\!\!\!0\,,\quad \varphi_2^+=0\,.
\end{eqnarray}
\end{subequations}
The field $H_g$ picks up the mass
\begin{equation}
m^2(H_g)=\frac 12 g^2 {w}^2\,,
\end{equation}
from the $D$-terms,
while the soft-SUSY breaking lifting terms give rise to the masses
\begin{equation}
m^2(H_0^R)= m^2(H_0^I) = m_1^2+m_2^2\,.
\end{equation}
Furthermore, from the kinetic terms in~(\ref{Lagrangian:U1}), we see that
$G_{\mathbbm{1}}$~(\ref{U1Goldstone}), which due to~(\ref{w2:semilocal:0})
is identical to
$G_{\tau^3}$, turns into the transverse component of $V_\mu$ according
to the Higgs mechanism.

This  setup has  similarities  to the  $F_D$-term hybrid  inflationary
scenario~\cite{Garbrecht:2006ft,Garbrecht:2006az}.     However,    the
mechanism used to  circumvent the gravitino problem in  that case does
not work here.  The Higgs state $H_g$, the ${\rm U}(1)_X$-vector boson
$A_\mu$  and  their  fermionic  superpartners,  all  attain  the  mass
$gw/\sqrt{2}$.  Because of  this, we have called all  these states the
$g$-sector  particles.  During  the  phase transition  at  the end  of
inflation,  ${w}$  is  evolving  rapidly,  such  that  the  $g$-sector
particles are produced {\it via} preheating, which occurs due to their
non-adiabatically  changing   mass  term.   Even   though  approximate
$D$-parity holds due to the smallness of~(\ref{Dviolation:semilocal}),
from~(\ref{X1X2})  and~(\ref{DMin}),   we  see  that   the  $g$-sector
particles can  undergo 3-body decays  into Goldstone bosons  and light
moduli. In contrast, this is  not possible when ${\rm SU}(2)$ is fully
gauged as in Refs.~\cite{Garbrecht:2006ft,Garbrecht:2006az}, where all
particles that have odd  $D$-parities acquire the mass $g{M}/\sqrt 2$,
such  that  these  types   of  decays  are  energetically  impossible.
However, we note  that in the semi-local case, we  can still resort to
the reheat scenarios outlined in Section~\ref{section:X:MSSM}.

Note that a possible mixing  between the gauged ${\rm U}(1)_X$ and the
${\rm U}(1)_Y$ of the SM  should be strongly suppressed.  Denoting the
coupling  of  $\widehat  X_{1,2}$  to  ${\rm  U}(1)_Y$-hypercharge  by
$g^X_Y$, we find the constraint
\begin{equation}
g^X_Y g \stackrel{<}{{}_\sim}
\frac{\Delta \vartheta_w}{\vartheta_w}\frac{v^2}{M^2}\,,
\end{equation}
where   $v$  is  the   VEV  of   the  SM   Higgs  boson   and  $\Delta
\vartheta_w/\vartheta_w\approx 7 \times 10^{-4}$ is the uncertainty in
Glashow's weak mixing angle $\vartheta_w$.

This     model     predicts     the    formation     of     semi-local
strings~\cite{Vachaspati:1991dz}.   Topology   does  not  protect  the
stability  of  semi-local strings.  When  they  unwind, the  potential
energy  is lowered  while there  is  an increase  in spatial  gradient
energy. If  the spatial gradient  energy is larger than  the potential
energy, the strings  are stable.  It turns out, that  this is the case
when~\cite{Hindmarsh:1991jq,Hindmarsh:1992yy,Achucarro:1992hs}
\begin{equation}
\frac{\kappa}{g}<\frac{1}{\sqrt{2}}\,,
\end{equation}
or equivalently $m^2(H^{R,I}_\kappa)<m^2(H_g)$.

The CMB  spectra for semi-local  strings will depend on  the parameter
$\kappa/g$.  It has  been speculated  that for  small  $\kappa/g$, the
spectrum  will  resemble that  of  Abelian  strings,  while for  large
values,    it     will    be     closer    to    that     of    global
textures~\cite{Achucarro:2007sp,Urrestilla:2007sf}.      A      recent
simulation  for the  case of  critical  coupling, $\kappa/g=1/\sqrt2$,
yields  indeed  temperature and  polarisation  spectra for  semi-local
strings that are  very similar, both in terms  of amplitude and shape,
to those  for textures~\cite{Urrestilla:2007sf}. One  will also expect
that  semi-local strings will  have a  large amount  of energy-density
stored  in scalar  field  gradients, which  is  leading to  long-range
interactions between  the strings. This  situation is very  similar to
the case  of textures, such that  non-Gaussian hot and  cold spots may
occur in the CMB.

\section{Inflation and Impact on the CMB}
\label{section:inflation}

We now discuss the predictions  of our model for the amplitude $A_{\rm
s}$ of the primordial power spectrum produced during inflation and the
scalar spectral index $n_{\rm s}$ that can be observed in the CMB.  As
we  discuss below,  for given  superpotential couplings,  the symmetry
breaking  scale  $M$  can  be  determined  by  imposing  the  measured
amplitude of the power spectrum.  In addition, the range of values for
$M$  that  explains   the  cold  spot  due  to   a  texture  is  given
by~\cite{Cruz:2007pe}
\begin{equation}
M=(6.2^{+1.5}_{-2.1})\times 10^{15}~{\rm GeV}\; .
\end{equation}
Notice that a factor of $1/\sqrt  2$ has to be considered here when we
compare  $M$  to  the  VEV~$\phi_0$ that  occurs  in~(\ref{V:texture})
and~(\ref{eta:obs}).    This   factor   arises  from   the   different
normalisations  used for  the  real fields  $\Phi_i$  and the  complex
fields $X_{1,2}$.  The fact that  the two determinations of $M$ should
be  compatible  enables  us  to  infer the  values  of  superpotential
couplings that lead to a successful explanation of the cold spot in an
inflation plus textures scenario. Given  the value of $M$, we can then
deduce the value of $n_{\rm s}$.

Our aim is now to show that our $F$-term hybrid inflationary model can
successfully  provide  quantitative  explanations  for both,  the  CMB
spectrum and  the cold  spot feature for  some values of  $\kappa$.  A
complete likelihood  analysis, yielding precision  measurements, would
require     an     exploration     of    parameter     space     using
Markov-Chain-Monte-Carlo techniques.  This  has been performed for the
case   of    SUSY   hybrid    inflation   and   cosmic    strings   in
Ref.~\cite{Battye:2006pk}.  Due to the similarity of the contributions
of  strings and  textures to  the CMB  spectrum, we  presume  that the
central   result   of   Ref.~\cite{Battye:2006pk},   that   a   string
contribution  close  to  the  maximum allowed  value  can  accommodate
$n_{\rm s}$ as large as 1, also applies to the case of textures.

Hybrid inflation  takes place when $\langle  S \rangle >  {M}$ and all
other  VEVs are  {\it  zero}. In  this  configuration, the  tree-level
potential~(\ref{V0:lambda:rho})    is    flat    in    the    inflaton
$\sigma=\sqrt2\,  {\rm Re}\,S$, such  that it  would not  roll towards
smaller   values.    However,  the   one-loop   contribution  to   the
inflationary      potential       is      (see,      for      example,
Ref.~\cite{Garbrecht:2006az})
\begin{eqnarray}
\label{V1loop}
\nonumber
V_{1-{\rm loop}} \!\!&=&\!\! \frac{1}{32\pi^2}\, \Bigg\{
2\kappa^4\, \Bigg[ |S^2 + {M}^2|^2\,
\ln\Bigg(\frac{\kappa^2 (|S|^2 +{M}^2)}{Q^2}\Bigg) +\,
|S^2 - {M}^2|^2\,
\ln\Bigg(\frac{\kappa^2 (|S|^2 - {M}^2)}{Q^2}\Bigg)\Bigg]\\
&&\hspace{-0.17cm} +\, 2\lambda^4\, \Bigg[
|S^2 + {\textstyle \frac{\kappa}{\lambda}}\, {M}^2|^2\,
\ln\Bigg(\frac{\lambda^2 (|S|^2 + {\textstyle \frac{\kappa}{\lambda}}
{M}^2)}{Q^2}\Bigg)\, +\,
|S^2 - {\textstyle \frac{\kappa}{\lambda}} {M}^2|^2\,
\ln\Bigg(\frac{\lambda^2 (|S|^2 - {\textstyle \frac{\kappa}{\lambda}}
{M}^2)}{Q^2}\Bigg)\Bigg]\nonumber\\
&&\hspace{-0.17cm} +\, \frac{3\rho^4}{2}\, \Bigg[
|S^2 + {\textstyle \frac{\kappa}{\rho}}\, {M}^2|^2\,
\ln\Bigg(\frac{\rho^2 (|S|^2 + {\textstyle \frac{\kappa}{\rho}}
{M}^2)}{Q^2}\Bigg)\, +\,
|S^2 - {\textstyle \frac{\kappa}{\rho}} {M}^2|^2\,
\ln\Bigg(\frac{\rho^2 (|S|^2 - {\textstyle \frac{\kappa}{\rho}}
{M}^2)}{Q^2}\Bigg)\Bigg]\nonumber\\
&&\hspace{-0.17cm} -\, |S|^4\, \Bigg[\, 2{\cal N}\kappa^4\,
\ln\Bigg(\frac{\kappa^2\,|S|^2}{Q^2}\Bigg)\: +\: 4\lambda^4\,
\ln\Bigg(\frac{\lambda^2\,|S|^2}{Q^2}\Bigg)
\: +\:
3\rho^4\,\ln\Bigg(\frac{\rho^2\, |S|^2}{Q^2}\Bigg)\,\Bigg]\,\Bigg\}
\; .
\end{eqnarray}
The  logarithmic slope  causes the  inflaton $\sigma$  to  slowly roll
towards smaller VEVs.  When $\langle S \rangle < {M}$, the combination
of waterfall fields,
\begin{equation}
\label{X:instability}
\frac{1}{\sqrt2} \left(X_1+X_2\right)\,,
\end{equation}
develops a  negative mass term which induces  spontaneous breakdown of
${\rm SU}(2)$.  In  order for this instability to  occur first in this
direction,  rather than  in any  other direction,  which  involves the
$N_i$  and $H_{u,d}$, we  either have  to impose  $\lambda>\kappa$ and
$\rho>\kappa$, or  we just decouple  these additional fields  from the
inflation sector  by setting $\lambda=0$  or $\rho=0$.  It is  at this
stage, when  the textures or  semi-local strings form.   The condition
$\lambda>\kappa$  can  be relaxed  if  a  flat  direction involving  a
top-squark has  a large  VEV during inflation  (see the  discussion in
Section~\ref{stop:VEV:delay}).  In  this case, $H_u$  acquires a large
mass which stabilises the VEVs of $H_{u,d}$, even if $\lambda<\kappa$.
In addition,  the mass splitting  between the $\widetilde  H_{u,d}$ is
suppressed,  such that in  the expression  for the  one-loop effective
potential,  we can  effectively set  $\lambda =0$,  regardless  of its
actual value.

In addition to the tree-level potential and the one-loop correction, we
also allow for a supergravity correction,
\begin{equation}
V_{\rm SUGRA}=32 \pi^2 \kappa^2 M^4 \frac{S^4}{m_{\rm Pl}^4}\,,
\end{equation}
where $m_{\rm Pl}=1.22\times 10^{19}~{\rm GeV}$ denotes the Planck mass.
For this expression, we have assumed that there are only renormalisable
contributions to the K\"ahler potential.

The  amplitude  of  the  power  spectrum at  the  scale  $k=0.002~{\rm
Mpc}^{-1}$ is given by
\begin{equation}
\label{PR}
A_{\rm s}=\frac{2^{7} \pi}{3 m_{\rm Pl}^6}
\frac{V^{3}(\sigma)}{(\partial V/\partial \sigma)^2}
\Bigg|_{\sigma=\sigma_{\rm e}}\,,
\end{equation}
where $\sigma_{\rm  e}$ is the value  of the inflaton  field when this
scale left the horizon. It can be determined from
\begin{equation}
\label{Ne}
N_{\rm e}=\frac{8\pi}{m_{Pl}^2}\int\limits_{\sqrt2 M}^{\sigma_{\rm e}}
d\sigma \left(\frac{\partial V}{\partial\sigma}\right)^{-1}\,,
\end{equation}
where $N_{\rm e}$ is the number of e-folds of inflation since horizon exit.
Here, we take $N_{\rm e}=55$, but note that there is a weak dependence
of this value on the thermal history of the Universe. This dependence has
no significant impact on the results presented in
this section and is therefore neglected.

Imposing the observed value $A_{\rm s}=2.35 \times 10^{-9}$, we can
determine the symmetry breaking scale $M$ for a given set of
superpotential couplings $\kappa$, $\rho$ and $\lambda$. In other
words, we determine the model parameter $M$ from the observed $A_{\rm
s}$.  The second observable we consider is the scalar spectral index
which is given by
\begin{equation}
n_{\rm s}=
1-2\frac{m_{\rm Pl}^2}{8\pi}
\frac{\partial^2V/\partial\sigma^2}{V}
\Bigg|_{\sigma=\sigma_{\rm e}}\,.
\end{equation}

\begin{figure}[t!]
\vskip.5cm
\begin{center}
\hskip-.2cm
\scalebox{1.}
{
\input{nmsugra}
}
\hfill
\hskip-.2cm
\scalebox{1.}
{
\input{nssugra}
}
\\
\scalebox{1.}
{
\input{nmnosugra}
}
\hfill
\hskip-.2cm
\scalebox{1.}
{
\input{nsnosugra}
}
\end{center}
\vskip-.5cm
\caption{
\label{figure:nmns}
\small The mass scale $M$ (left panels) and the scalar spectral index
$n_{\rm s}$ (right panels) against $\kappa$ for the hybrid inflationary
models considered in the text. In the upper two panels, we include
supergravity corrections, whereas in the bottom two panels, we
do not inlude supergravity corrections. 
The solid lines correspond to
$\lambda=\rho=0$ and the dotted lines to $\lambda=0$,
$\rho=\kappa$. In the left
panels, the solid horizontal lines indicate the
best-fitting symmetry breaking mass scale required for textures to
account for the observed cold spot. The $1\sigma$ lower and upper
bounds are also displayed by thin dashed lines.
}
\end{figure}
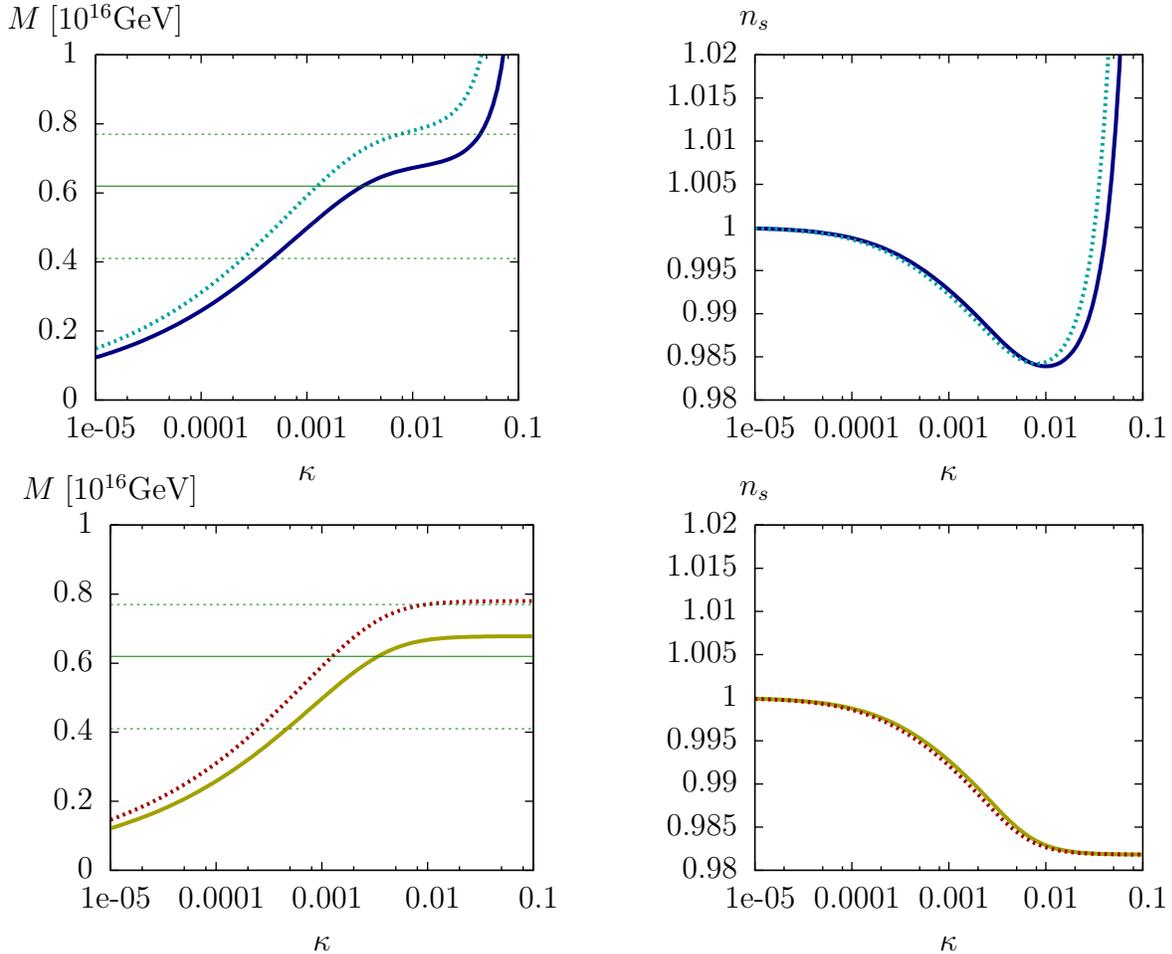

We can  now compare the  predictions of our $F$-term  hybrid inflation
model    with    the    best-fit    value    for    $M$    found    in
Ref.~\cite{Cruz:2007pe}.      The    result     is     presented    in
Fig.~\ref{figure:nmns}.  We plot the prediction for $M$ imposed by the
normalisation  of $A_{\rm s}$  for a  model with  $\lambda=\rho=0$ and
another model with $\lambda=0$ and $\rho=\kappa$, according to the two
different reheat  scenarios discussed in Section~\ref{section:X:MSSM}.
For  both  cases,  we  also   show  the  effect  of  the  supergravity
correction, by comparing to the case where this correction is included
to  the case  where it  is omitted.  To guide  the eye,  we  have also
included   the   best-fit   value   for   the   texture   scale   from
Ref.~\cite{Cruz:2007pe}, along with the  $1\sigma$ bounds. We see that
for the models  we consider, a satisfactory amplitude  for the texture
perturbation       can       be       achieved      for       $2\times
10^{-4}\stackrel{<}{{}_\sim}\kappa\stackrel{<}{{}_\sim}5\times
10^{-2}$.


We  also  plot  the  prediction  for  the  scalar  spectral  index  in
Fig.~\ref{figure:nmns}. For the values of $\kappa$ which would explain
the cold spot, the scalar spectral index $n_{\rm s}$ lies roughly in a
range  between  $0.985$  and  $1.01$.  If there  were  only  adiabatic
perturbations, these high values would  be in tension with the present
data~\cite{Dunkley:2008ie}, which requires a smaller value for $n_{\rm
s}$.  However,  in case  there is a  sub-dominant contribution  to the
CMB-spectrum from  topological defects,  larger values of  $n_{\rm s}$
are consistent  with the data if  the defect contribution  is close to
its maximum allowed value. Since textures have a similar impact on the
CMB to strings, we conjecture that a texture contribution close to its
maximum allowed  value will  also allow for  larger values  of $n_{\rm
s}$.

Indeed,  the maximum  allowed value  for  the texture  scale from  CMB
observations  is~\cite{Urrestilla:2007sf} $M\leq7.4\times 10^{15}~{\rm
GeV}$,  such that  the  the value  suggested by~\cite{Cruz:2007pe}  is
close to  maximal.  
Consequently, the  high values of  $n_{\rm s}$ indicated  by our
analysis  do not necessarily  constitute a problem
for the texture model at present.

\section{Conclusions}

The current CMB data shows  hints for the presence  of topological
defects of order the GUT  energy scale $M_{\rm GUT} \sim 10^{16}$~GeV.
First, fits to  the CMB spectrum marginally favour  models with a 10\%
contribution to  it that  results from defects.   Second, it  has been
suggested that textures associated  with GUT energies could provide an
explanation    of    the    origin    of   the    non-Gaussian    cold
spot~\cite{Cruz:2007pe}.  In the near future, improved measurements of
the  CMB  spectrum on  small  scales,  for  example by  {\it  Planck},
ACT~\cite{Huffenberger:2004gm,Kosowsky:2006na}                      and
SPT~\cite{Ruhl:2004kv},  will either  severely tighten  the  bounds on
defects or strongly hint to their presence~\cite{Pogosian:2008am}. For
instance,  detailed  estimates of  the  bounds  on  the cosmic  string
tension that are  anticipated to be possible using  the results of the
Planck satellite are given in Ref.~\cite{Battye:2007si}.

In this paper  we have presented a realistic  model of $F$-term hybrid
inflation which can lead to the formation of textures at the waterfall
transition.  The  main results  of  our  study  may be  summarised  as
follows:

\begin{itemize}

\item  The   superpotential  of   the  $F$-term  hybrid   model  under
  consideration  realises a  ${\rm  SU}(2)\times{\rm SU}(2)_X$  global
  symmetry  which  is  explicitly  broken  to  ${\rm  SU}(2)\times{\rm
  U}(1)_X$  by soft-SUSY  breaking terms.   The global  symmetry ${\rm
  SU}(2)\times{\rm U}(1)_X$ is spontaneously broken to ${\rm U(1)}_D$,
  giving  rise   to  a  vacuum   manifold  that  is   homeomorphic  to
  $S^3$.  Thus,  the model  can  naturally  predict  the formation  of
  textures  at  the   waterfall  transition.   If  the  superpotential
  coupling    $\kappa$     is    not    too     small,    i.e.~$\kappa
  \stackrel{>}{{}_\sim}  10^{-4}$, the  energy scale  of  the produced
  textures  will  be  close  to   the  GUT  scale,  as  was  suggested
  in~\cite{Cruz:2007pe}.

\item The  aforementioned $F$-term hybrid  model contains a  number of
  flat  directions that  are lifted  by the  small  soft-SUSY breaking
  masses.   The presence of  these quasi-flat  directions needs  to be
  taken  carefully into  account in  the reheating  process.   We have
  exemplified  a  viable   cosmological  scenario,  incorporating  the
  transfer of  energy from the inflationary to  the MSSM sector, while
  avoiding gravitino overproduction or fast stellar cooling.

\item  We  have  also  presented  a number  of  alternative  scenarios
  including a semi-local model, where  $U(1)_{\rm X}$ is promoted to a
  local  symmetry.  Detailed  discussion of  explicit  and spontaneous
  breaking  patterns  due  to   the  presence  of  the  moduli-lifting
  soft SUSY-breaking  terms are given in the  appendices.  Our results
  may  also   be  useful  for  building  other   models  with  related
  symmetries.

\end{itemize}

In conclusion, the analysis presented in this paper has shown that the
symmetry  breaking scale  $\phi_0\approx 8.7\times  10^{15}~{\rm GeV}$
required   to   fit   the   cold   spot  can   be   achieved   without
field-theoretical, phenomenological  or cosmological difficulties. The
$F$-term  hybrid model  under study  provides a  predictive, realistic
framework  that  can  successfully  address the  recent  evidence  for
texture  defects present in  the current  CMB data  and link  these to
inflation.

\section*{Acknowledgements}

The work of BG was supported  in part by the U.S. Department of Energy
contracts  No.  DE-FG02-08ER41531  and  DE-FG02-95ER40896 and  by  the
Wisconsin Alumni Research Foundation.  The work of AP was supported in
part by the STFC research grant: PP/D000157/1.

\newpage

\begin{appendix}

\section{Symmetries and Breaking Patterns}
\label{Appendix:ExtraSymmetry}
\renewcommand{\theequation}{\ref{Appendix:ExtraSymmetry}.\arabic{equation}}
\setcounter{equation}{0}

In this  appendix, we present  more details on the  ${\rm SU}(2)\times
{\rm  SU}(2)_X$ symmetry  of  the superpotential.   In particular,  we
discuss formal  aspects of the spontaneous and  the explicit breakdown
of  the symmetry,  such as  counting of  Goldstone modes  and homotopy
groups of the vacuum manifold.

The  combined  action   of  the  symmetry  transformations~(\ref{SU2})
and~(\ref{SU2X}) can be compactly written in terms of a tensor-product
as
\begin{equation}
\label{symtrafo:tensor}
\mathbf{X}=
\left(
\begin{array}{c}
\widehat X_1\\
{\rm i}\tau^2 \widehat X_2
\end{array}
\right)\
\xrightarrow{{\rm SU}(2)_{\rm X}\times{\rm SU}(2)}\
{\rm e}^{{\rm i}\theta_j \sigma^j} \otimes {\rm e}^{{\rm i}\alpha_i \tau^i}
\left(
\begin{array}{c}
\widehat X_1\\
{\rm i}\tau^2 \widehat X_2
\end{array}
\right)\,.
\end{equation}
To be more explicit, the action of ${\rm SU}(2)_{\rm X}$ only is given by
\begin{equation}
\left(
\begin{array}{c}
\widehat X_1^+\\
\widehat X_1^-\\
\widehat X_2^-\\
-\widehat X_2^+\\
\end{array}
\right)\
\xrightarrow{{\rm SU}(2)_{\rm X}}\
\exp\left[{\rm i}\theta_i
\left(
\begin{array}{cccc}
\sigma^i_{11} & 0 & \sigma^i_{12} & 0\\
0 & \sigma^i_{11} & 0 & \sigma^i_{12}\\
\sigma^i_{21} & 0 & \sigma^i_{22} & 0\\
0 & \sigma^i_{21} & 0 & \sigma^i_{22}\\
\end{array}
\right)
\right]
\left(
\begin{array}{c}
\widehat X_1^+\\
\widehat X_1^-\\
\widehat X_2^-\\
-\widehat X_2^+\\
\end{array}
\right)
\,.
\end{equation}
As  an  immediate   application  of  this,  we  may   write  down  the
superpotential~(\ref{superpotential:SU2})  in   the  manifestly  ${\rm
SU}(2)\times{\rm SU}(2)_X$-invariant form:
\begin{equation}
\widehat X_1^T \widehat X_2\ =\
-\ \frac{1}{2}\; 
{\mathbf X}^T {\rm i}\sigma^2 \otimes {\rm i} \tau^2\, {\mathbf X}\; .
\end{equation}
Since we have 6 symmetries available, and the scalar fields $X_1$ and
$X_2$ comprise 8 degrees of freedom, we can always reparameterise
the VEVs to have the form
\begin{equation}
\label{parametrization:six:symmetries}
X_1=
\left(\begin{array}{c}
{w}_1^+\\
0
\end{array}
\right)
\,,\quad
X_2=
\left(\begin{array}{c}
{w}_2^+\\
0
\end{array}
\right)
\,.
\end{equation}
We continue to denote the ${\rm SU}(2)_X$ generators by $\sigma^i$
and those of ${\rm SU}(2)$ by $\tau^i$.


We first discuss the symmetry breaking in the case that there is no
additional explicit breaking through the lifting potential $V_{\rm
lift}$ given by~(\ref{Vlift}).  Using the parameterisation of the
VEVs~(\ref{parametrization:six:symmetries}), it is useful to
distinguish two cases. An overview of arguments is presented in
Table~\ref{Table:SSB:noexplicitSB}.
\begin{itemize}
\item
When $w_1^+=w_2^+$ ($|X_1|=|X_2|$ is the reparameterization
invariant form), all 3 generators of ${\rm SU}(2)$
and ${\rm SU}(2)_X$ have the same effect on the VEVs, which
means $G_{\tau^i}=G_{\sigma^i}$.
Due to this identity, the symmetry is broken to a diagonal group,
according to
\begin{equation}
\label{pattern:X1eqX2}
{\rm SU}(2)\times {\rm SU}(2)_X\xrightarrow{|X_1|=|X_2|}
{\rm SU}(2)_D\,.
\end{equation}
In  this case,  the additional  massless modes  $H_0^{R,I}$  and $H_g$
cannot  be identified  as Goldstone  bosons.  For  the  third homotopy
group associated with the  symmetry breaking, we find $\pi_3\left({\rm
SU}(2)\times  {\rm SU}(2)_X/{\rm  SU}(2)_D\right)= \mathbbm  Z$, while
all other homotopy groups are  equal to the identity. Therefore, there
are topologically  stable textures and,  up to the  non-Goldstone flat
directions, the  vacuum manifold is homeomorphic  to the 3-dimensional
sphere $S^3$.
\item
When       $w_1^+\not=w_2^+$       (reparameterization      invariant:
$|X_1|\not=|X_2|$), we only  have $G_{\tau^3}=G_{\sigma^3}$, such that
only a diagonal Abelian symmetry remains intact according to
\begin{equation}
\label{pattern:X1neqX2}
{\rm SU}(2)\times {\rm SU}(2)_X\xrightarrow{|X_1|\not=|X_2|}
{\rm U}(1)_D\,.
\end{equation}
The  5 Goldstone  bosons  are the  $G_{\tau^i}$  and $H_0^{I,R}$.  The
latter   two   are  linear   combinations   of  $G_{\tau^{1,2}}$   and
$G_{\sigma^{1,2}}$  which  are  orthogonal  to  $G_{\tau^{1,2}}$.  The
remaining  flat  direction  $H_g$  corresponds to  the  transformation
$w_1^+\to  \zeta  w_1^+$, $w_2^+\to  \zeta^{-1}  w_2^+$, which  leaves
$w_1^+  w_2^+=M^2$ invariant.  The  third homotopy  group is  given by
$\pi_3\left({\rm   SU}(2)\times  {\rm   SU}(2)_X/{\rm  U}(1)_D\right)=
\mathbbm  Z$  and the  second  by  $\pi_2\left({\rm SU}(2)\times  {\rm
SU}(2)_X/{\rm  U}(1)_D\right)= \mathbbm  Z$,  such that  there may  be
monopoles  besides  textures. Indeed,  up  to  the non-Goldstone  flat
direction,  the vacuum  manifold is  homeomorphic to  $S^3\times S^2$.
The emergence of the $S^2$ has a simple geometric interpretation.  For
fixed  values of $|X_1|$  and $|X_2|$,  the condition  $F_S=0$ implies
that $X_1  X_2 = {M}^2$, which  fixes the absolute value  of the angle
between $X_1$ and  $X_2$, when considering each as  a four dimensional
real  vector. Hence,  keeping $|X_1|$  and $|X_2|$  fixed,  the vacuum
manifold is a direct product  of the 3-dimensional sphere $S^3$, times
an $S^2$  corresponding to rotations that leave  this angle invariant.
When taking  the flat direction $H_g$ into  account, the configuration
can   change   to  $|X_1|=|X_2|$,   such   that   the  monopoles   can
unwind.  Hence, the  monopoles  are topologically  unstable. In  other
words,  even though  $\pi_2\left({\rm SU}(2)\times  {\rm SU}(2)_X/{\rm
U}(1)_D\right)= \mathbbm  Z$, for the full vacuum  manifold ${\cal M}$
including the  direction $H_g$, we find $\pi_2({\cal  M})=0$.  We have
not  checked whether  the monopoles  will nevertheless  be dynamically
stable, that  is, whether  there is a  barrier in the  gradient energy
which prevents them from unwinding.
\end{itemize}

\begin{table}
\begin{tabular}{|l|l|l|l|l|}
\hline
Configuration & SSB &
\begin{tabular}{l}
Goldstone\\
modes
\end{tabular}
&
\begin{tabular}{l}
Vacuum\\
manifold
\end{tabular}
& 
\begin{tabular}{l}
Topological\\
defects
\end{tabular}
\\
\hline
\vspace{-.45cm}
&&&&\\
\hline
$|X_1|=|X_2|$ &
$
\begin{array}{l}
{\rm SU}(2)\times {\rm SU}(2)_X\\
\xrightarrow{|X_1|=|X_2|}{\rm SU}(2)_D
\end{array}
$
& $G_{\tau^i}$ ($i=1,2,3$) &
\begin{tabular}{l}
$S^3$ (up to\\
non-Goldstone\\
flat directions)
\end{tabular}
& textures\\
\hline
$|X_1|\not=|X_2|$ &
$
\begin{array}{l}
{\rm SU}(2)\times {\rm SU}(2)_X\\
\xrightarrow{|X_1|\not=|X_2|}{\rm U}(1)_D
\end{array}
$
&
\begin{tabular}{l}
$G_{\tau^i}$ ($i=1,2,3$),\\
$H_0^I$, $H_0^R$\\
\end{tabular}
&
\begin{tabular}{l}
$S^3\times S^2$ (up to\\
non-Goldstone\\
flat direction)
\end{tabular}
&
\begin{tabular}{l}
textures and\\
topologically\\
unstable\\
monopoles
\end{tabular}
\\
\hline
\end{tabular}
\caption{
\label{Table:SSB:noexplicitSB}
Symmetry breaking patterns for $V_{\rm lift} = 0$.}
\end{table}

We now turn to the case $V_{\rm lift}\not=0$, which is relevant for
the main body of the paper. Again, we distinguish two situations and 
a summary of these arguments is presented in
Table~\ref{Table:SSB:explicitSB}.
\begin{itemize}
\item
Suppose  first that in  (\ref{Vlift}), we  have $m_1=m_2$.   Since the
transformation~(\ref{symtrafo:tensor})     leaves    $|X_1|^2+|X_2|^2$
invariant  and  hence $V_{\rm  lift}$  is  also  invariant.  For  this
reason, (\ref{dda})  explicitly gives $|X_1|=|X_2|$,  and the symmetry
breaking  pattern  is   as  in~(\ref{pattern:X1eqX2}).   There  are  3
Goldstone  modes $G_{\tau^i}$  ($i=1,2,3$)  and the  moduli $H_g$  and
$H_0^{R,I}$ acquire masses as in (\ref{moduli:masses}), since they are
not  associated with  a broken  symmetry.   In this  case, the  vacuum
manifold is an $S^3$ and there may be texture defects.
\item
If  $m_1\not=m_2$,  $V_{\rm  lift}$   is  no  longer  invariant  under
transformations   generated    by   the   ${\rm   SU}(2)_X$-generators
$\sigma^{1,2}$.    Moreover,    we   find   from    (\ref{dda})   that
$|X_1|\not=|X_2|$.   Since  $V_{\rm lift}$  is  still invariant  under
rotations  related to  the ${\rm  SU}(2)_X$-generator  $\sigma^3$, the
symmetry breaking pattern is
\begin{equation}
{\rm SU}(2)\times{\rm SU}(2)_X
\xrightarrow{V_{\rm lift}}
{\rm SU}(2)\times{\rm U}(1)_X
\xrightarrow{|X_1|\not=|X_2|}
{\rm U}(1)_D\,.
\end{equation}
In this situation, there are again 3 Goldstone bosons $G_{\tau^i}$
($i=1,2,3$).  The moduli  $H_0^{R,I}$  can become  massive, since  the
breaking  effect of  $V_{\rm  lift}$ turns  them  from Goldstone  into
pseudo-Goldstone bosons  of ${\rm SU}(2)_X$. The  remaining mode $H_g$
is not associated with a broken symmetry and can, therefore, acquire a
mass. The vacuum manifold is  again $S^3$, and accordingly, there will
be textures.
\end{itemize}

\begin{table}
\begin{tabular}{|l|l|l|l|l|}
\hline
\begin{tabular}{l}
Lifting\\
terms
\end{tabular}
&
\begin{tabular}{l}
Breaking\\
pattern
\end{tabular}
&
\begin{tabular}{l}
Goldstone\\
modes
\end{tabular}
&
\begin{tabular}{l}
Pseudo-\\
Goldstone\\
modes
\end{tabular}
& 
\begin{tabular}{l}
Vacuum\\
manifold\\
and defects
\end{tabular}
\\
\hline
\vspace{-.45cm}
&&&&\\
\hline
$m_1=m_2$ &
$
\begin{array}{l}
{\rm SU}(2)\times {\rm SU}(2)_X\\
\xrightarrow{\mbox{\footnotesize explicit}}\\
{\rm SU}(2)\times {\rm SU}(2)_X\\
\xrightarrow{|X_1|=|X_2|} {\rm SU}(2)_D
\end{array}
$
& $G_{\tau^i}$ ($i=1,2,3$) &
none
& 
\begin{tabular}{l}
$S^3$,\\
textures
\end{tabular}
\\
\hline
$m_1 \not = m_2$ &
$
\begin{array}{l}
{\rm SU}(2)\times {\rm SU}(2)_X\\
\xrightarrow{\mbox{\footnotesize explicit}}\\
{\rm SU}(2)\times {\rm U}(1)_X\\
\xrightarrow{|X_1|\not=|X_2|} {\rm U}(1)_D
\end{array}

$
&
\begin{tabular}{l}
$G_{\tau^i}$ ($i=1,2,3$)
\end{tabular}
&
\begin{tabular}{l}
$H_0^I$, $H_0^R$
\end{tabular}
&
\begin{tabular}{l}
$S^3$,\\
textures
\end{tabular}
\\
\hline
\end{tabular}
\caption{
\label{Table:SSB:explicitSB}
Symmetry breaking  patterns for $V_{\rm  lift} \neq 0$. In  the second
column of the  table, we also exhibit the  symmetry breaking steps for
the  cases $m_1=m_2$  and $m_1  \neq m_2$  due to  the  {\em explicit}
presence of a non-zero $V_{\rm lift}$.}
\end{table}

Finally, we  comment on the  semi-local model described  in Section~4.
In this case, the ${\rm U}(1)_X$ is singled out of ${\rm SU}(2)_X$ due
to  the  gauging,  which  induces  gauge  kinetic  terms  as  well  as
$D$-terms.   These  terms   explicitly  break  ${\rm  SU}(2)_X$.   The
symmetry breaking pattern is
\begin{equation}
{\rm SU}(2)\times{\rm SU}(2)_X
\xrightarrow{\textnormal{gauge}}
{\rm SU}(2)\times{\rm U}(1)_X
\xrightarrow{|X_{1,2}|\not=0}
{\rm U}(1)_D\,.
\end{equation}
Again, there  are 3 Goldstone bosons $G_{\tau^i}$  ($i=1,2,3$), out of
which $G_{\tau^3}$  is absorbed as  a transverse degree of  freedom of
the  ${\rm U}(1)_X$-gauge  boson. The  vacuum manifold  is  $S^3$, and
there will be semi-local strings.

\section{Field Redefinitions and the Lifting Potential}
\label{Appendix:FieldRedef}
\renewcommand{\theequation}{\ref{Appendix:FieldRedef}.\arabic{equation}}
\setcounter{equation}{0}

When we include the $d$-term in the lifting potential~(\ref{Vlift}),
it turns out that the minima~(\ref{dda}) get shifted to
\begin{subequations}
\label{VEVs:dterm}
\begin{eqnarray}
\label{dda:dterm}
{w}_1^+\!\!\!&=&\!\!\!\sqrt{\frac{m_2}{m_1}{M}^2
  \frac{1}{\sqrt{1-\frac{|d|^2}{m_1^2 m_2^2}}}-
  \frac{m_2^2}{\kappa^2}} 
\,,\\ 
{w}_2^+\!\!\!&=&\!\!\!\sqrt{1-\frac{|d|^2}{m_1^2 m_2^2}}
\sqrt{\frac{m_1}{m_2}{M}^2 \frac{1}{\sqrt{1-\frac{|d|^2}{m_1^2
  m_2^2}}}- \frac{m_1^2}{\kappa^2}} 
\,,\\
{w}_2^-\!\!\!&=&\!\!\!\frac{|d|}{m_2}
\sqrt{\frac{{M}^2}{m_1 m_2} \frac{1}{\sqrt{1-\frac{|d|^2}{m_1^2
  m_2^2}}}- \frac{1}{\kappa^2}} 
\,,\\
\label{phi2m}
\alpha_3 -\beta \!\!\!&=&\!\!\!\pi+\arg d
\,,
\end{eqnarray}
\end{subequations}
and the moduli masses are modified to be
\begin{eqnarray}
\label{mmass:dterm}
m^2(H_0^R) = m^2(H_0^I) = m_1^2+m_2^2\,,\quad m^2(H_g)= 4 \frac{m_1^2
  m_2^2-|d|^2}{m_1^2 + m_2^2}\,. 
\end{eqnarray}
For a given $\alpha_3$, $\beta$ is now fixed by
(\ref{phi2m}), which is due to the fact that
the $d$-term explicitly breaks the ${\rm U}(1)_X$-symmetry. Therefore,
$\beta$ does not correspond to a massless mode. We recall
from (\ref{dda}), that
in the situation where $d=0$, $m_1\not=0$ and $m_2\not=0$,
it follows that $w_2^-=0$ and that the parameters $\alpha_3$
and $\beta$ induce the same rotations. Hence, in both situations,
$\beta$ does not correspond to an independent
massless degree of freedom.

This is consistent with the fact that when performing a field
redefinition, the $d$-term can be removed.  To see this, we make use
of the fact that field redefinitions according to ${\rm SU}(2)_X$ have
the explicit effect
\begin{subequations}
\begin{eqnarray}
\hspace{-.8cm}
|X_1|^2\!\!\!&\xrightarrow{\exp({\rm i}\theta_1 \sigma^1)}&\!\!\!
|X_1|^2 \cos^2\theta_1 + |X_2|^2 \sin^2\theta_1
+{\rm i}\sin\theta_1\cos\theta_1
\left(
X_1^\dagger {\rm i}\tau^2 X_2 + X_2^\dagger{\rm i}\tau^2 X_1
\right)\,,\\
\hspace{-.8cm}
|X_2|^2\!\!\!&\xrightarrow{\exp({\rm i}\theta_1 \sigma^1)}&\!\!\!
|X_2|^2 \cos^2\theta_1 + |X_1|^2 \sin^2\theta_1
-{\rm i}\sin\theta_1\cos\theta_1
\left(
X_2^\dagger {\rm i}\tau^2 X_1 + X_1^\dagger{\rm i}\tau^2 X_2
\right)\,,
\end{eqnarray}
\end{subequations}
while a  rotation by  $\exp({\rm i}\vartheta_3 \sigma^3)$  changes the
phase  of  $d$ in  the  lifting  potential.\footnote{  Note that  this
particular symmetry is identical to the ${\rm U}(1)_X$.}

Hence, we find that, upon ${\rm SU}(2)_{\rm X}$ field redefinitions,
any $d$-term can be absorbed into the $m_{1,2}^2$-terms, provided
$|d|^2<m_1^2 m_2^2$. If the latter condition is not fulfilled,
we see from~(\ref{mmass:dterm}), that the mode $H_g$ attains a negative
mass square and the potential has no stable minimum.
We also note that the moduli masses in~(\ref{mmass:dterm}) are invariant
under this type of reparameterization, as they should.

\section{Scalar Interaction Terms and $D$-Parities}
\label{Appendix:DParities}
\renewcommand{\theequation}{\ref{Appendix:DParities}.\arabic{equation}}
\setcounter{equation}{0}

In   this   appendix,    we   present   the   expressions~(\ref{X1X2})
and~(\ref{DMin}), from which the interactions within the scalar sector
can  easily be inferred.   It turns  out that  in the  situation where
${w}_1^+  \approx {w}_2^+$,  the symmetry  of the  model  is enhanced,
which   has   the    consequence   that   certain   interactions   are
suppressed.  These   accidental  symmetries  are   important  for  the
$F_D$-model,   and  we   adapt   the  term   $D$-parities  from   that
context~\cite{Garbrecht:2006az}.

In the global model, for $m_1=m_2$,
$V=V_0+V_{\rm lift}$ given by~(\ref{V0}) and~(\ref{Vlift}) is invariant under
the discrete accidental symmetries
\begin{eqnarray}
D_1:&&\qquad X_1 \leftrightarrow X_2
\,,\\
D_2:&&\qquad X_1 \leftrightarrow \tau^3 X_1\,,
\qquad
X_2 \leftrightarrow \tau^3 X_2\,.
\end{eqnarray}
For the semi-local case, we have to require $m_{\rm FI}=0$ in
addition, for the $D$ parities to be respected.  In
Table~\ref{table:Dparities}, we list the $D$-parities of the
particular mass eigenstates. If $m_1=m_2$, interactions that violate
the discrete multiplicative $D$-parities are forbidden. On the other
hand, allowing for $m_1\not=m_2$ or, respectively, $m_{\rm FI}\not=0$,
the smallness of ${{w}_1^+}^2-{{w}_2^+}^2$ allows to control the rate
of $D$-parity violating interactions, a feature which we make use of
when constructing the semi-local model.

\begin{table}
\begin{center}
\begin{tabular}{|c||c|c|}
\hline
Eigenstate & $\,D_1\mbox{-parity}\!\!$
& $\,D_2 \textnormal{-parity}\!\!$
\\
\hline\vspace{-1.6mm}&\vspace{-1.6mm}&\vspace{-1.6mm}\\\hline
$H_\kappa^R\,$,
$H_\kappa^I$
&
$+$
&
$+$
\\
\hline
$G_{\tau^1}\,$,
$H_0^R$
&
$-$
&
$-$
\\
\hline
$G_{\tau^2}\,$,
$H_0^I$
&
$+$
&
$-$
\\
\hline
$G_{\tau^3}\,$,
$H_g$
&
$-$
&
$+$
\\
\hline
\end{tabular}
\end{center}
\caption{\small $D$-parities of the scalar mass
  eigenstates. \label{table:Dparities}} 
\end{table}

In particular,  we explicitly see  the realisation of  $D$-parity when
writing down the expression occurring within the $F_S$-term,
\begin{eqnarray}
\label{X1X2}
X_1 X_2\!\!\!&=&\!\!\!{w}_1^+{w}_2^+
+\frac{{w}}{\sqrt2}(H_\kappa^R+{\rm i}H_\kappa^I)
\\
\nonumber
&&\!\!\!+\frac{{w}_1^+{w}_2^+}{2{w}^2}\left(
G_{\tau^1}^2+G_{\tau^2}^2+G_{\tau^3}^2-H_g^2-2 {\rm i}H_g G_{\tau^3}
-{H_0^R}^2-{H_0^I}^2
+(H_\kappa^R +{\rm i} H_\kappa^I)^2
\right)\\
\nonumber
&&\!\!\!-\frac{\rm i}{2}G_{\tau^1}H_0^R-\frac{\rm i}{2}G_{\tau^2}H_0^I
\\
\nonumber
&&\!\!\!
+\frac12\frac{{{w}_1^+}^2-{{w}_2^+}^2}{{{w}_1^+}^2+{{w}_2^+}^2}
\left(
G_{\tau^2} H_0^R -G_{\tau^1} H_0^I -G_{\tau^3}({\rm i}H_\kappa^R -
H_\kappa^I) + H_g H_\kappa^R +{\rm i} H_g H_\kappa^I 
\right)\,, 
\end{eqnarray}
where the last term is the $D$-parity violating contribution.
Accordingly, the $D$-term~(\ref{DTerm}) is
\begin{eqnarray}
\label{DMin}
D\!\!\!&=&\!\!\!-\frac{g}{2}m_{\rm FI}^2 +
\frac{g}{2}
\frac{{{w}_1^+}^4-{{w}_2^+}^4}{{{w}_1^+}^2+{{w}_2^+}^2}
\\
\nonumber
&&
+\frac{g}{\sqrt{2}}{w} H_g
+g\frac{{w}_1^+ {w}_2^+}{{{w}_1^+}^2+{{w}_2^+}^2}
\left(
H_g H_\kappa^R + H_0^I G_{\tau^1} - H_0^R G_{\tau^2} +H_\kappa^I G_{\tau^3}
\right)
\\
\nonumber
&&
+\frac{g}{4{w}^2}
\frac{{{w}_1^+}^2-{{w}_2^+}^2}{{{w}_1^+}^2+{{w}_2^+}^2}\left(
-{H_\kappa^R}^2-{H_\kappa^I}^2-{H_0^R}^2-{H_0^I}^2+H_g^2
+G_{\tau^1}^2+G_{\tau^2}^2+G_{\tau^3}^2
\right)\,.
\end{eqnarray}
Note  that  the  third  and   fourth  terms  are  both  $D_1$-odd  and
$D_2$-even,   while  the   remaining   terms  are   even  under   both
parities. Hence, for $m_{1,2}=0$  and $m_{\rm FI}=0$, the $D$-term has
definite $D$-parity, so $D^2$ conserves $D$-parity.

Since  we assume $m_{1,2}\ll  M$, we  may parameterise  the $D$-parity
violation in the global model by
\begin{equation}
\label{Dviolation:global}
\frac{{{w}_1^+}^2-{{w}_2^+}^2}{{{w}_1^+}^2+{{w}_2^+}^2}
\approx\frac{m_2^2-m_1^2}{2 m_1 m_2}\,,
\end{equation}
where we have set $d=0$ and used~(\ref{dda}).
In the semi-local model, using~(\ref{VEV:semilocal}), we find
\begin{equation}
\label{Dviolation:semilocal}
\frac{{{w}_1^+}^2-{{w}_2^+}^2}{{{w}_1^+}^2+{{w}_2^+}^2}
\approx \frac{m_{\rm FI}^2}{2{M}^2}+\frac{m_2^2-m_1^2}{g^2 {M}^2}\,.
\end{equation}
Hence, we  note that  $D$-parity violation in  the semi-local  case is
always  suppressed  as   $m_{1,2}^2/{M}^2$  or  $m_{\rm  FI}^2/{M}^2$,
whereas the corresponding expression~(\ref{Dviolation:global}) for the
global case can be of order one in general, unless $m_1\approx m_2$.

\end{appendix}

\end{document}

%% file: nmsugra.tex
\begingroup
  \makeatletter
  \providecommand\color[2][]{%
    \GenericError{(gnuplot) \space\space\space\@spaces}{%
      Package color not loaded in conjunction with
      terminal option `colourtext'%
    }{See the gnuplot documentation for explanation.%
    }{Either use 'blacktext' in gnuplot or load the package
      color.sty in LaTeX.}%
    \renewcommand\color[2][]{}%
  }%
  \providecommand\includegraphics[2][]{%
    \GenericError{(gnuplot) \space\space\space\@spaces}{%
      Package graphicx or graphics not loaded%
    }{See the gnuplot documentation for explanation.%
    }{The gnuplot epslatex terminal needs graphicx.sty or graphics.sty.}%
    \renewcommand\includegraphics[2][]{}%
  }%
  \providecommand\rotatebox[2]{#2}%
  \@ifundefined{ifGPcolor}{%
    \newif\ifGPcolor
    \GPcolortrue
  }{}%
  \@ifundefined{ifGPblacktext}{%
    \newif\ifGPblacktext
    \GPblacktexttrue
  }{}%
  \let\gplgaddtomacro\g@addto@macro
  \gdef\gplbacktext{}%
  \gdef\gplfronttext{}%
  \makeatother
  \ifGPblacktext
    \def\colorrgb#1{}%
    \def\colorgray#1{}%
  \else
    \ifGPcolor
      \def\colorrgb#1{\color[rgb]{#1}}%
      \def\colorgray#1{\color[gray]{#1}}%
      \expandafter\def\csname LTw\endcsname{\color{white}}%
      \expandafter\def\csname LTb\endcsname{\color{black}}%
      \expandafter\def\csname LTa\endcsname{\color{black}}%
      \expandafter\def\csname LT0\endcsname{\color[rgb]{1,0,0}}%
      \expandafter\def\csname LT1\endcsname{\color[rgb]{0,1,0}}%
      \expandafter\def\csname LT2\endcsname{\color[rgb]{0,0,1}}%
      \expandafter\def\csname LT3\endcsname{\color[rgb]{1,0,1}}%
      \expandafter\def\csname LT4\endcsname{\color[rgb]{0,1,1}}%
      \expandafter\def\csname LT5\endcsname{\color[rgb]{1,1,0}}%
      \expandafter\def\csname LT6\endcsname{\color[rgb]{0,0,0}}%
      \expandafter\def\csname LT7\endcsname{\color[rgb]{1,0.3,0}}%
      \expandafter\def\csname LT8\endcsname{\color[rgb]{0.5,0.5,0.5}}%
    \else
      \def\colorrgb#1{\color{black}}%
      \def\colorgray#1{\color[gray]{#1}}%
      \expandafter\def\csname LTw\endcsname{\color{white}}%
      \expandafter\def\csname LTb\endcsname{\color{black}}%
      \expandafter\def\csname LTa\endcsname{\color{black}}%
      \expandafter\def\csname LT0\endcsname{\color{black}}%
      \expandafter\def\csname LT1\endcsname{\color{black}}%
      \expandafter\def\csname LT2\endcsname{\color{black}}%
      \expandafter\def\csname LT3\endcsname{\color{black}}%
      \expandafter\def\csname LT4\endcsname{\color{black}}%
      \expandafter\def\csname LT5\endcsname{\color{black}}%
      \expandafter\def\csname LT6\endcsname{\color{black}}%
      \expandafter\def\csname LT7\endcsname{\color{black}}%
      \expandafter\def\csname LT8\endcsname{\color{black}}%
    \fi
  \fi
  \setlength{\unitlength}{0.0500bp}%
  \begin{picture}(4464.00,3528.00)%
    \gplgaddtomacro\gplbacktext{%
      \csname LTb\endcsname%
      \put(770,660){\makebox(0,0)[r]{\strut{} 0}}%
      \put(770,1181){\makebox(0,0)[r]{\strut{} 0.2}}%
      \put(770,1702){\makebox(0,0)[r]{\strut{} 0.4}}%
      \put(770,2222){\makebox(0,0)[r]{\strut{} 0.6}}%
      \put(770,2743){\makebox(0,0)[r]{\strut{} 0.8}}%
      \put(770,3264){\makebox(0,0)[r]{\strut{} 1}}%
      \put(902,440){\makebox(0,0){\strut{} 1e-05}}%
      \put(1699,440){\makebox(0,0){\strut{} 0.0001}}%
      \put(2496,440){\makebox(0,0){\strut{} 0.001}}%
      \put(3293,440){\makebox(0,0){\strut{} 0.01}}%
      \put(4090,440){\makebox(0,0){\strut{} 0.1}}%
      \put(2496,110){\makebox(0,0){\strut{}$\kappa$}}%
      \put(902,3524){\makebox(0,0){$M\;[{10^{16}\rm GeV}]$}}%
    }%
    \gplgaddtomacro\gplfronttext{%
    }%
    \gplbacktext
    \put(0,0){\includegraphics{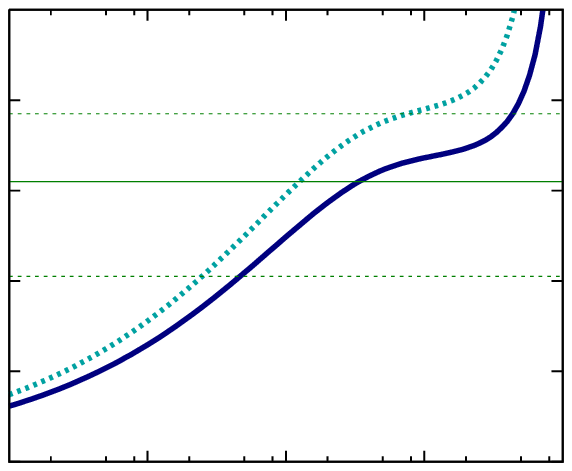}}%
    \gplfronttext
  \end{picture}%
\endgroup

%% file: nssugra.tex
\begingroup
  \makeatletter
  \providecommand\color[2][]{%
    \GenericError{(gnuplot) \space\space\space\@spaces}{%
      Package color not loaded in conjunction with
      terminal option `colourtext'%
    }{See the gnuplot documentation for explanation.%
    }{Either use 'blacktext' in gnuplot or load the package
      color.sty in LaTeX.}%
    \renewcommand\color[2][]{}%
  }%
  \providecommand\includegraphics[2][]{%
    \GenericError{(gnuplot) \space\space\space\@spaces}{%
      Package graphicx or graphics not loaded%
    }{See the gnuplot documentation for explanation.%
    }{The gnuplot epslatex terminal needs graphicx.sty or graphics.sty.}%
    \renewcommand\includegraphics[2][]{}%
  }%
  \providecommand\rotatebox[2]{#2}%
  \@ifundefined{ifGPcolor}{%
    \newif\ifGPcolor
    \GPcolortrue
  }{}%
  \@ifundefined{ifGPblacktext}{%
    \newif\ifGPblacktext
    \GPblacktexttrue
  }{}%
  \let\gplgaddtomacro\g@addto@macro
  \gdef\gplbacktext{}%
  \gdef\gplfronttext{}%
  \makeatother
  \ifGPblacktext
    \def\colorrgb#1{}%
    \def\colorgray#1{}%
  \else
    \ifGPcolor
      \def\colorrgb#1{\color[rgb]{#1}}%
      \def\colorgray#1{\color[gray]{#1}}%
      \expandafter\def\csname LTw\endcsname{\color{white}}%
      \expandafter\def\csname LTb\endcsname{\color{black}}%
      \expandafter\def\csname LTa\endcsname{\color{black}}%
      \expandafter\def\csname LT0\endcsname{\color[rgb]{1,0,0}}%
      \expandafter\def\csname LT1\endcsname{\color[rgb]{0,1,0}}%
      \expandafter\def\csname LT2\endcsname{\color[rgb]{0,0,1}}%
      \expandafter\def\csname LT3\endcsname{\color[rgb]{1,0,1}}%
      \expandafter\def\csname LT4\endcsname{\color[rgb]{0,1,1}}%
      \expandafter\def\csname LT5\endcsname{\color[rgb]{1,1,0}}%
      \expandafter\def\csname LT6\endcsname{\color[rgb]{0,0,0}}%
      \expandafter\def\csname LT7\endcsname{\color[rgb]{1,0.3,0}}%
      \expandafter\def\csname LT8\endcsname{\color[rgb]{0.5,0.5,0.5}}%
    \else
      \def\colorrgb#1{\color{black}}%
      \def\colorgray#1{\color[gray]{#1}}%
      \expandafter\def\csname LTw\endcsname{\color{white}}%
      \expandafter\def\csname LTb\endcsname{\color{black}}%
      \expandafter\def\csname LTa\endcsname{\color{black}}%
      \expandafter\def\csname LT0\endcsname{\color{black}}%
      \expandafter\def\csname LT1\endcsname{\color{black}}%
      \expandafter\def\csname LT2\endcsname{\color{black}}%
      \expandafter\def\csname LT3\endcsname{\color{black}}%
      \expandafter\def\csname LT4\endcsname{\color{black}}%
      \expandafter\def\csname LT5\endcsname{\color{black}}%
      \expandafter\def\csname LT6\endcsname{\color{black}}%
      \expandafter\def\csname LT7\endcsname{\color{black}}%
      \expandafter\def\csname LT8\endcsname{\color{black}}%
    \fi
  \fi
  \setlength{\unitlength}{0.0500bp}%
  \begin{picture}(4464.00,3528.00)%
    \gplgaddtomacro\gplbacktext{%
      \csname LTb\endcsname%
      \put(1034,660){\makebox(0,0)[r]{\strut{} 0.98}}%
      \put(1034,986){\makebox(0,0)[r]{\strut{} 0.985}}%
      \put(1034,1311){\makebox(0,0)[r]{\strut{} 0.99}}%
      \put(1034,1637){\makebox(0,0)[r]{\strut{} 0.995}}%
      \put(1034,1962){\makebox(0,0)[r]{\strut{} 1}}%
      \put(1034,2287){\makebox(0,0)[r]{\strut{} 1.005}}%
      \put(1034,2613){\makebox(0,0)[r]{\strut{} 1.01}}%
      \put(1034,2938){\makebox(0,0)[r]{\strut{} 1.015}}%
      \put(1034,3264){\makebox(0,0)[r]{\strut{} 1.02}}%
      \put(1166,440){\makebox(0,0){\strut{} 1e-05}}%
      \put(1897,440){\makebox(0,0){\strut{} 0.0001}}%
      \put(2628,440){\makebox(0,0){\strut{} 0.001}}%
      \put(3359,440){\makebox(0,0){\strut{} 0.01}}%
      \put(4090,440){\makebox(0,0){\strut{} 0.1}}%
      \put(2628,110){\makebox(0,0){\strut{}$\kappa$}}%
      \put(1166,3524){\makebox(0,0){$n_s$}}%
    }%
    \gplgaddtomacro\gplfronttext{%
    }%
    \gplbacktext
    \put(0,0){\includegraphics{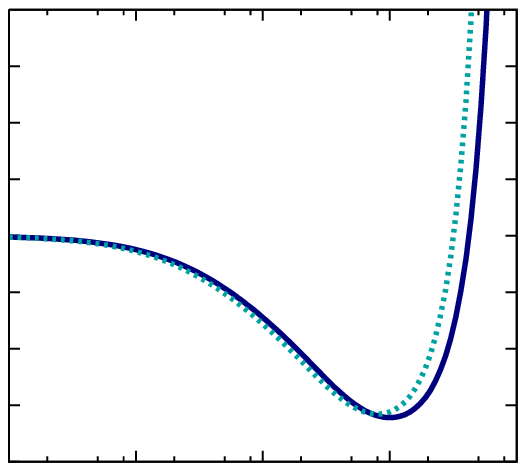}}%
    \gplfronttext
  \end{picture}%
\endgroup

%% file: nmnosugra.tex
\begingroup
  \makeatletter
  \providecommand\color[2][]{%
    \GenericError{(gnuplot) \space\space\space\@spaces}{%
      Package color not loaded in conjunction with
      terminal option `colourtext'%
    }{See the gnuplot documentation for explanation.%
    }{Either use 'blacktext' in gnuplot or load the package
      color.sty in LaTeX.}%
    \renewcommand\color[2][]{}%
  }%
  \providecommand\includegraphics[2][]{%
    \GenericError{(gnuplot) \space\space\space\@spaces}{%
      Package graphicx or graphics not loaded%
    }{See the gnuplot documentation for explanation.%
    }{The gnuplot epslatex terminal needs graphicx.sty or graphics.sty.}%
    \renewcommand\includegraphics[2][]{}%
  }%
  \providecommand\rotatebox[2]{#2}%
  \@ifundefined{ifGPcolor}{%
    \newif\ifGPcolor
    \GPcolortrue
  }{}%
  \@ifundefined{ifGPblacktext}{%
    \newif\ifGPblacktext
    \GPblacktexttrue
  }{}%
  \let\gplgaddtomacro\g@addto@macro
  \gdef\gplbacktext{}%
  \gdef\gplfronttext{}%
  \makeatother
  \ifGPblacktext
    \def\colorrgb#1{}%
    \def\colorgray#1{}%
  \else
    \ifGPcolor
      \def\colorrgb#1{\color[rgb]{#1}}%
      \def\colorgray#1{\color[gray]{#1}}%
      \expandafter\def\csname LTw\endcsname{\color{white}}%
      \expandafter\def\csname LTb\endcsname{\color{black}}%
      \expandafter\def\csname LTa\endcsname{\color{black}}%
      \expandafter\def\csname LT0\endcsname{\color[rgb]{1,0,0}}%
      \expandafter\def\csname LT1\endcsname{\color[rgb]{0,1,0}}%
      \expandafter\def\csname LT2\endcsname{\color[rgb]{0,0,1}}%
      \expandafter\def\csname LT3\endcsname{\color[rgb]{1,0,1}}%
      \expandafter\def\csname LT4\endcsname{\color[rgb]{0,1,1}}%
      \expandafter\def\csname LT5\endcsname{\color[rgb]{1,1,0}}%
      \expandafter\def\csname LT6\endcsname{\color[rgb]{0,0,0}}%
      \expandafter\def\csname LT7\endcsname{\color[rgb]{1,0.3,0}}%
      \expandafter\def\csname LT8\endcsname{\color[rgb]{0.5,0.5,0.5}}%
    \else
      \def\colorrgb#1{\color{black}}%
      \def\colorgray#1{\color[gray]{#1}}%
      \expandafter\def\csname LTw\endcsname{\color{white}}%
      \expandafter\def\csname LTb\endcsname{\color{black}}%
      \expandafter\def\csname LTa\endcsname{\color{black}}%
      \expandafter\def\csname LT0\endcsname{\color{black}}%
      \expandafter\def\csname LT1\endcsname{\color{black}}%
      \expandafter\def\csname LT2\endcsname{\color{black}}%
      \expandafter\def\csname LT3\endcsname{\color{black}}%
      \expandafter\def\csname LT4\endcsname{\color{black}}%
      \expandafter\def\csname LT5\endcsname{\color{black}}%
      \expandafter\def\csname LT6\endcsname{\color{black}}%
      \expandafter\def\csname LT7\endcsname{\color{black}}%
      \expandafter\def\csname LT8\endcsname{\color{black}}%
    \fi
  \fi
  \setlength{\unitlength}{0.0500bp}%
  \begin{picture}(4464.00,3528.00)%
    \gplgaddtomacro\gplbacktext{%
      \csname LTb\endcsname%
      \put(770,660){\makebox(0,0)[r]{\strut{} 0}}%
      \put(770,1181){\makebox(0,0)[r]{\strut{} 0.2}}%
      \put(770,1702){\makebox(0,0)[r]{\strut{} 0.4}}%
      \put(770,2222){\makebox(0,0)[r]{\strut{} 0.6}}%
      \put(770,2743){\makebox(0,0)[r]{\strut{} 0.8}}%
      \put(770,3264){\makebox(0,0)[r]{\strut{} 1}}%
      \put(902,440){\makebox(0,0){\strut{} 1e-05}}%
      \put(1699,440){\makebox(0,0){\strut{} 0.0001}}%
      \put(2496,440){\makebox(0,0){\strut{} 0.001}}%
      \put(3293,440){\makebox(0,0){\strut{} 0.01}}%
      \put(4090,440){\makebox(0,0){\strut{} 0.1}}%
      \put(2496,110){\makebox(0,0){\strut{}$\kappa$}}%
      \put(902,3524){\makebox(0,0){$M\;[{10^{16}\rm GeV}]$}}%
    }%
    \gplgaddtomacro\gplfronttext{%
    }%
    \gplbacktext
    \put(0,0){\includegraphics{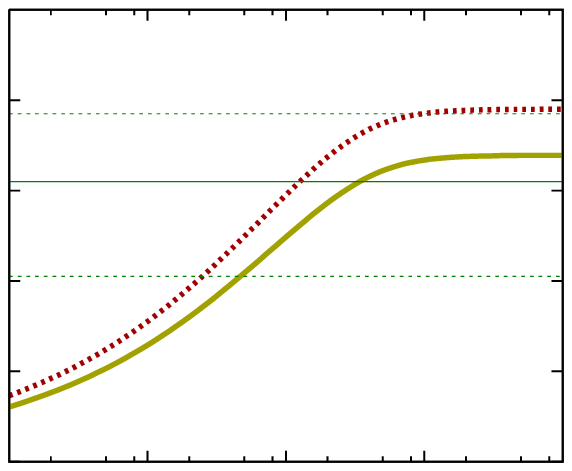}}%
    \gplfronttext
  \end{picture}%
\endgroup

%% file: nsnosugra.tex
\begingroup
  \makeatletter
  \providecommand\color[2][]{%
    \GenericError{(gnuplot) \space\space\space\@spaces}{%
      Package color not loaded in conjunction with
      terminal option `colourtext'%
    }{See the gnuplot documentation for explanation.%
    }{Either use 'blacktext' in gnuplot or load the package
      color.sty in LaTeX.}%
    \renewcommand\color[2][]{}%
  }%
  \providecommand\includegraphics[2][]{%
    \GenericError{(gnuplot) \space\space\space\@spaces}{%
      Package graphicx or graphics not loaded%
    }{See the gnuplot documentation for explanation.%
    }{The gnuplot epslatex terminal needs graphicx.sty or graphics.sty.}%
    \renewcommand\includegraphics[2][]{}%
  }%
  \providecommand\rotatebox[2]{#2}%
  \@ifundefined{ifGPcolor}{%
    \newif\ifGPcolor
    \GPcolortrue
  }{}%
  \@ifundefined{ifGPblacktext}{%
    \newif\ifGPblacktext
    \GPblacktexttrue
  }{}%
  \let\gplgaddtomacro\g@addto@macro
  \gdef\gplbacktext{}%
  \gdef\gplfronttext{}%
  \makeatother
  \ifGPblacktext
    \def\colorrgb#1{}%
    \def\colorgray#1{}%
  \else
    \ifGPcolor
      \def\colorrgb#1{\color[rgb]{#1}}%
      \def\colorgray#1{\color[gray]{#1}}%
      \expandafter\def\csname LTw\endcsname{\color{white}}%
      \expandafter\def\csname LTb\endcsname{\color{black}}%
      \expandafter\def\csname LTa\endcsname{\color{black}}%
      \expandafter\def\csname LT0\endcsname{\color[rgb]{1,0,0}}%
      \expandafter\def\csname LT1\endcsname{\color[rgb]{0,1,0}}%
      \expandafter\def\csname LT2\endcsname{\color[rgb]{0,0,1}}%
      \expandafter\def\csname LT3\endcsname{\color[rgb]{1,0,1}}%
      \expandafter\def\csname LT4\endcsname{\color[rgb]{0,1,1}}%
      \expandafter\def\csname LT5\endcsname{\color[rgb]{1,1,0}}%
      \expandafter\def\csname LT6\endcsname{\color[rgb]{0,0,0}}%
      \expandafter\def\csname LT7\endcsname{\color[rgb]{1,0.3,0}}%
      \expandafter\def\csname LT8\endcsname{\color[rgb]{0.5,0.5,0.5}}%
    \else
      \def\colorrgb#1{\color{black}}%
      \def\colorgray#1{\color[gray]{#1}}%
      \expandafter\def\csname LTw\endcsname{\color{white}}%
      \expandafter\def\csname LTb\endcsname{\color{black}}%
      \expandafter\def\csname LTa\endcsname{\color{black}}%
      \expandafter\def\csname LT0\endcsname{\color{black}}%
      \expandafter\def\csname LT1\endcsname{\color{black}}%
      \expandafter\def\csname LT2\endcsname{\color{black}}%
      \expandafter\def\csname LT3\endcsname{\color{black}}%
      \expandafter\def\csname LT4\endcsname{\color{black}}%
      \expandafter\def\csname LT5\endcsname{\color{black}}%
      \expandafter\def\csname LT6\endcsname{\color{black}}%
      \expandafter\def\csname LT7\endcsname{\color{black}}%
      \expandafter\def\csname LT8\endcsname{\color{black}}%
    \fi
  \fi
  \setlength{\unitlength}{0.0500bp}%
  \begin{picture}(4464.00,3528.00)%
    \gplgaddtomacro\gplbacktext{%
      \csname LTb\endcsname%
      \put(1034,660){\makebox(0,0)[r]{\strut{} 0.98}}%
      \put(1034,986){\makebox(0,0)[r]{\strut{} 0.985}}%
      \put(1034,1311){\makebox(0,0)[r]{\strut{} 0.99}}%
      \put(1034,1637){\makebox(0,0)[r]{\strut{} 0.995}}%
      \put(1034,1962){\makebox(0,0)[r]{\strut{} 1}}%
      \put(1034,2287){\makebox(0,0)[r]{\strut{} 1.005}}%
      \put(1034,2613){\makebox(0,0)[r]{\strut{} 1.01}}%
      \put(1034,2938){\makebox(0,0)[r]{\strut{} 1.015}}%
      \put(1034,3264){\makebox(0,0)[r]{\strut{} 1.02}}%
      \put(1166,440){\makebox(0,0){\strut{} 1e-05}}%
      \put(1897,440){\makebox(0,0){\strut{} 0.0001}}%
      \put(2628,440){\makebox(0,0){\strut{} 0.001}}%
      \put(3359,440){\makebox(0,0){\strut{} 0.01}}%
      \put(4090,440){\makebox(0,0){\strut{} 0.1}}%
      \put(2628,110){\makebox(0,0){\strut{}$\kappa$}}%
      \put(1166,3524){\makebox(0,0){$n_s$}}%
    }%
    \gplgaddtomacro\gplfronttext{%
    }%
    \gplbacktext
    \put(0,0){\includegraphics{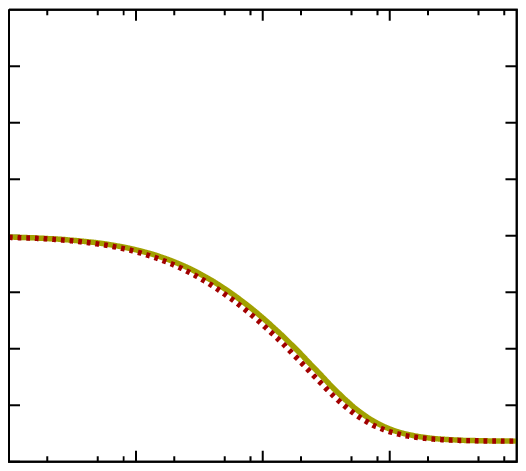}}%
    \gplfronttext
  \end{picture}%
\endgroup